\documentclass[12pt,draftcls,journal,leterpaper,twosides,onecolumn]{IEEEtran}

\usepackage[english]{babel}
\usepackage{amsmath}		
\usepackage{graphics}		
\usepackage{caption}
\usepackage{subcaption}
\usepackage{color,psfrag}
\usepackage{pstool}
\usepackage{setspace}		
\usepackage{longtable}          
\usepackage{amssymb,amscd,latexsym,dsfont}
\usepackage{float,color,graphicx,balance}
\usepackage{multirow,multicol}
\usepackage{comment,cite}
\usepackage{enumerate}
\usepackage{balance}
\usepackage{stfloats}
\usepackage{tikz}
\usetikzlibrary{fit}					
\usetikzlibrary{backgrounds}	
\usetikzlibrary{matrix}
\usetikzlibrary{calc}
\usepackage{array,arydshln}
\usepackage{cite}
\usepackage{mathtools}
\usepackage{algpseudocode}
\usepackage[linesnumbered,ruled]{algorithm2e}

\definecolor{refkey}{rgb}{1,0.5,0} 
\definecolor{labelkey}{rgb}{1,0.5,0}

\usepackage{glossaries}
\glossarystyle{listdotted}
\makeglossaries
\renewcommand*{\CustomAcronymFields}{%
  name={\the\glsshorttok},
  description={\the\glslongtok},
  first={\noexpand\emph{\the\glslongtok}\space(\the\glsshorttok)},%
  firstplural={\noexpand\emph{\the\glslongtok\noexpand\acrpluralsuffix}\space(\the\glsshorttok)},%
  text={\the\glsshorttok},%
  plural={\the\glsshorttok\noexpand\acrpluralsuffix}%
}

\def \scalevalueS {0.9}
\def \scalevalueSS {0.87}

\DeclareMathOperator\erfc{erfc}

\newcommand{\tr}[1]{\textrm{#1}}
\newcommand{\ttr}[1]{\textrm{\footnotesize{#1}}}

\newcommand{\PR}[1]{\Pr\left\{#1\right\}}       

\newcommand{\secref}[1]{Sec.~\ref{#1}}

\newcommand{\figref}[1]{Fig.~\ref{#1}}

\newcommand{\tabref}[1]{Table~\ref{#1}}
\newcommand{\Algref}[1]{Alg.~\ref{#1}}

\newcommand{\Pout}{P_{\tr{\tiny{out}}}}
\newcommand{\Nbar}{\bar{N}}
\newcommand{\Tbar}{\bar{T}}

\newcommand{\tA}[2]{A_{#1}^{#2}}				
\newcommand{\rA}[2]{\tilde{A}_{#1}^{#2}}		
\newcommand{\tC}[1]{C_{#1}}						
\newcommand{\rC}[1]{\tilde{C}_{#1}}				
\newcommand{\pn}{p_0}							
\newcommand{\pnb}{\bar{p_0}}					
\newcommand{\pa}{p_1}							
\newcommand{\pab}{\bar{p_1}}					
\newcommand{\pe}[1]{p_{e}^{#1}}			
\newcommand{\pkt}[2]{\mathsf{P}_{#1}^{#2}}					
\newcommand{\rpkt}[2]{\tilde{\mathsf{P}}_{#1}^{#2}}	
\newcommand{\qn}{q_0}							
\newcommand{\qnb}{\bar{\qn}}					
\newcommand{\qa}{q_1}							
\newcommand{\qab}{\bar{\qa}}					
\newcommand{\ef}[2]{F_{#1}^{#2}}					
\newcommand{\es}[2]{S_{#1}^{#2}}					

\allowdisplaybreaks

\begin{document}

\newacronym{csi}{CSI}{channel state information}
\newacronym{cqi}{CQI}{channel quality indicator}
\newacronym{ack}{ACK}{acknowledgement}
\newacronym{arq}{ARQ}{automatic repeat request}
\newacronym{awgn}{AWGN}{additive white Gaussian noise}
\newacronym{cc}{CC}{chase combining}
\newacronym{dp}{DP}{dynamic programming}
\newacronym{fec}{FEC}{forward error correction}
\newacronym{harq}{HARQ}{hybrid automatic repeat request}
\newacronym{hspa}{HSPA}{high speed packet access}
\newacronym{iid}{i.i.d.}{independent and identically distributed}
\newacronym{ir}{IR}{incremental redundancy}
\newacronym{lte}{LTE}{long term evolution}
\newacronym{mdp}{MDP}{markov decision process}
\newacronym{mrc}{MRC}{maximal-ratio combining}
\newacronym{nack}{NAK}{negative acknowledgement}
\newacronym{pdf}{pdf}{probability density function}
\newacronym{wimax}{WiMax}{worldwide interoperability for microwave access}
\newacronym{3gpp}{3GPP}{3rd generation partnership project}
\newacronym{ofdm}{OFDM}{orthogonal frequency-division multiplexing}
\newacronym{ofdma}{OFDMA}{orthogonal frequency-division multiple access}
\newacronym{wlan}{WLAN}{wireless local area network}
\newacronym{gsm}{GSM}{global system for mobile communications}
\newacronym{edge}{EDGE}{enhanced data \gls{gsm} environment}
\newacronym{amc}{AMC}{adaptive modulation and coding}
\newacronym{snr}{SNR}{signal to noise ratio}
\newacronym{sinr}{SINR}{signal to interference and noise ratio}
\newacronym{mi}{MI}{mutual information}
\newacronym{acmi}{ACMI}{accumulated mutual information}
\newacronym{nacmi}{NACMI}{normalized ACMI}
\newacronym{cdi}{CDI}{channel distribution information}
\newacronym{latr}{LATR}{long-term average transmission rate}
\newacronym{rtr}{RTR}{round transmission rate}
\newacronym{pomdp}{POMDP}{Partially Observable Markov Decision Process}
\newacronym{fd}{FD}{full-duplex}
\newacronym{hd}{HD}{half-duplex}
\newacronym{td}{TD}{Time Division}
\newacronym{tdma}{TDMA}{time division multiple access}
\newacronym{mac}{MAC}{Media Access Control}
\newacronym{uwb}{UWB}{Ultra Wideband}
\newacronym{ieee}{IEEE}{institute of electrical and electronics engineers}
\newacronym{dB}{dB}{decibel}
\newacronym{cdf}{cdf}{cumulative density function}
\newacronym{ccdf}{ccdf}{complementary cumulative density function}
\newacronym{min}{Min.}{minimum}
\newacronym{med}{Med.}{median}
\newacronym{avg}{Avg.}{average}
\newacronym{ul}{UL}{up-link}
\newacronym{dl}{DL}{down-link}
\newacronym{app}{APP}{a-posteriori probability}
\newacronym{logmap}{LogMAP}{log maximum a-posteriori}
\newacronym{llr}{LLR}{log-likelihood ratio}
\newacronym{ue}{UE}{user equipment}
\newacronym{qos}{QoS}{quality of service}
\newacronym{5g}{5G}{5\textsuperscript{th} generation mobile networks}
\newacronym{4g}{4G}{4\textsuperscript{th} generation mobile networks}
\newacronym{tti}{TTI}{transmission time interval}
\newacronym{rrm}{RRM}{radio resource management}
\newacronym{mmib}{MMIB}{mean mutual information per bit}
\newacronym{dsi}{DSI}{decoder state information}
\newacronym{tb}{TB}{transport block}
\newacronym{tbs}{TBS}{transport block size}
\newacronym{cb}{CB}{code block}
\newacronym{cbg}{CBG}{code block group}
\newacronym{cbs}{CBS}{code block size}
\newacronym{prb}{PRB}{physical resource block}
\newacronym{rb}{RB}{resource block}
\newacronym{bler}{BLER}{block error rate}
\newacronym{blep}{BLEP}{block error probability}
\newacronym{crc}{CRC}{cyclic redundancy check}
\newacronym{tdd}{TDD}{time division duplex}
\newacronym{fdd}{FDD}{frequency division duplex}
\newacronym{embb}{eMBB}{enhanced mobile broadband}
\newacronym{mcc}{MCC}{mission critical communication}
\newacronym{mmc}{MMC}{massive machine communication}
\newacronym{mtc}{MTC}{machine type of communication}
\newacronym{mmtc}{mMTC}{massive machine type of communication}
\newacronym{umtc}{uMTC}{ultra-reliable \gls{mtc}}
\newacronym{urllc}{URLLC}{ultra-reliable low latency communication}
\newacronym{rtt}{RTT}{round trip time}
\newacronym{rs}{RS}{reference symbols}
\newacronym{kpi}{KPI}{key performance indicator}
\newacronym{kpis}{KPIs}{key performance indicators}
\newacronym{tx}{Tx}{transmitter node}
\newacronym{rx}{Rx}{receiver node}
\newacronym{cran}{C-RAN}{centralized radio access network}
\newacronym{rru}{RRU}{remote radio unit}
\newacronym{bbu}{BBU}{baseband unit}
\newacronym{fhd}{FHD}{fronthaul delay}
\newacronym{cch}{CCH}{control channel}
\newacronym{saw}{SAW}{stop-and-wait}
\newacronym{qci}{QCI}{\gls{qos} class identifier}
\newacronym{gbr}{GBR}{guaranteed bit rate}
\newacronym{mbr}{MBR}{maximum bit rate}
\newacronym{ngbr}{non-GBR}{non-\gls{gbr}}
\newacronym{arp}{ARP}{allocation and retention priority}
\newacronym{effcr}{ECR}{effective coding rate}
\newacronym{cbs}{CBS}{code block size}
\newacronym{mcs}{MCS}{modulation and coding scheme}
\newacronym{eva}{EVA}{extended vehicular A}
\newacronym{epa}{EPA}{extended pedestrian A}
\newacronym{etu}{ETU}{extended typical urban}
\newacronym{re}{RE}{resource element}
\newacronym{reS}{REs}{resource elements}
\newacronym{nr}{NR}{new radio}
\newacronym{qpsk}{QPSK}{quadrature phase shift keying}
\newacronym{qam}{QAM}{quadrature amplitude modulation}
\newacronym{siso}{SISO}{single-input and single-output}
\newacronym{bs}{BS}{base station}
\newacronym{phy}{PHY}{physical layer}
\newacronym{rlc}{RLC}{radio link control}
\newacronym{bcfsaw}{BCF-SAW}{BCF-SAW}
\newacronym{bcf}{BCF}{backwards composite feedback}
\newacronym{bac}{BAC}{binary asymmetric channel}
\newacronym{bsc}{BSC}{binary symmetric channel}
\newacronym{dtx}{DTX}{discontinued transmission}
\newacronym{bpsk}{BPSK}{binary phase shift keying}
\newacronym{bep}{BEP}{bit error probability}
\newacronym{ndi}{NDI}{new data indicator}
\newacronym{regsaw}{Reg-SAW}{Regular SAW}
\newacronym{Lrep}{$L$-Rep-ACK}{Increased feedback repetition order}
\newacronym{Lack}{$L$-ACK-SAW}{$L$ required ACK per packet}
\newacronym{RetxL}{ReTx-$L$-ACK}{Retransmission until $L$ ACKs are observed}
\newacronym{Asym}{Asym-SAW}{Asymmetric feedback detection for SAW}
\newacronym{bretx}{Blind-ReTx}{Blind retransmission}
\newacronym{dci}{DCI}{downlink control information}

\bstctlcite{IEEEexample:BSTcontrol}

\title{Analysis of Feedback Error in Automatic Repeat reQuest}

\author{
    \IEEEauthorblockN{Saeed R. Khosravirad and Harish Viswanathan \\ Nokia - Bell Labs}
\vspace{-40pt}
}

\maketitle

\begin{abstract}
The future  wireless networks envision ultra-reliable communication with efficient use of the limited wireless channel resources. Closed-loop  repetition protocols where retransmission  of a packet is enabled using a feedback channel has been adopted since early days of wireless telecommunication. Protocols such as \gls{arq} are used in today's wireless technologies as a mean to provide the link with reduced rate of packet outage and increased average throughput.  The performance of such protocols is strongly dependent to the feedback channel reliability. This paper studies the problem of feedback error and proposes a new method of acknowledging packet delivery for  retransmission protocols in unreliable feedback channel conditions. The proposed method is based on backwards composite acknowledgment from multiple packets in a retransmission protocol and  provides the scheduler of the wireless channel with additional parameters  to configure ultra-reliable communication for a user depending on channel quality. Numerical analysis are presented which show orders of magnitude increase in reliability of the proposed method as compared to \gls{arq} at the cost of a small  increase in average experienced delay.
\end{abstract}

\IEEEpeerreviewmaketitle

\section{Introduction}

Repetition of a packet  over non-deterministic channel conditions is a prominent approach to  reliable  packet delivery. Wireless telecommunications technologies such as \gls{hspa}, \gls{wimax} and \gls{lte}, to mention a few, have  relied on the performance boost provided by  retransmission techniques such as \gls{arq} and \gls{harq} \cite{Lin:1984}. Such retransmission protocols  add to the robustness of  transmission and increase link throughput. In \gls{lte}, and as expected for the \gls{5g} \cite{3gpp38802}, \gls{arq} is used in the \gls{rlc} layer while \gls{harq} in the lower \gls{mac} and upper \gls{phy} layer. Performing together, these retransmission protocols   provide the system with  high reliability where failure in the \gls{mac} layer \gls{harq} operation  is  compensated for by the \gls{rlc} layer \gls{arq} in acknowledged mode at the expense of extra  experienced latency  for the packet \cite{3gpp36212}.

The role of feedback channel is to limit repetitions  to only when the initial attempt is failed thus, increasing data channel efficiency. However, inevitable feedback channel impairments   may cause unreliability in packet delivery.  A  decoding failure report, i.e. \gls{nack}, falsely received as positive  \gls{ack}  results in undesirable packet outage.  Attempts to increase feedback reliability, e.g., by means of repetition coding, is costful to the receiver node while erronoeus feedback detection may cause an increased packet delivery latency and diminish  throughput and reliability \gls{kpis}. E.g., in  \gls{lte}   a single-bit \gls{ack}/\gls{nack}  spans over multiple \gls{re} up to a  \gls{prb} in  \gls{ul} and \gls{dl} \gls{harq} respectively to reduce false feedback detection \cite{3gpp36213}, making  feedback bits too costly to the network. In newer releases of \gls{lte}, blind \gls{harq} retransmissions of a packet is considered as a solution to avoid feedback complexity of broadcast \gls{harq} and increase reliability \cite{3gpp36877}. Such approach, despite the offered simplicity, can severely decrease resource utilization efficiency of the system considering that typically a high percentage  of  transmissions are successfully decoded in the initial attempt in typical link adaptation configurations. 

The core question this paper tries to answer is how to reliably  design a feedback-based retransmission protocol in unreliabile feedback conditions. We first study the effect of erroneous feedback  on performance of retransmission protocols.  We assume a simple \gls{saw} mode of operation in a narrow-band wireless link where the receiver node is a low-cost and low-energy device with limited power for feedback channel acknowledgement reports. Such model portrays well the unreliable feedback channel problem where the straight-forward solution to acquire reliable packet delivery is by either adding  diversity gain to the feedback link or relaxing the dependency to feedback channel and performing blind or conservative retransmission of the packet. Specifically, for low-cost  narrow-band communication such diversity gain can be achieved by increasing time diversity order of the feedback channel. We study different approaches of increasing feedback channel time diversity and  establish achievable reliability regions with respect to feedback channel error rate. We  show that  in reasonably reliable feedback channel conditions where the product of packet error rate and feedback error rate is comparable to packet outage rate, conservative asymmetric feedback detection can provide the required reliability level by slightly increasing false \gls{nack} rate while reducing false \gls{ack} rate. Further, in extremely un-reliable feedback channel conditions we see that blind retransmission of packet is the  viable solution in terms of reliability while it zeros the receiver node's energy consumption over feedback channel.

Next, we propose a new method of backwards composite acknowledgment that helps improves  reliability of repetition process without the need to increase time diversity order of the feedback channel.   The proposed scheme  relies on collaboration between  transmitter and receiver nodes to provide ultra-reliable communication of packets even in poor feedback channel conditions. Furthermore, thanks to the additional design parameters provided by  the proposed method, the scheduler of wireless network is able to configure each communication node with desirable ultra-reliability  only using one layer of retransmission protocol. This  enables the wireless technologies such as \gls{lte} to adopt one layer of retransmission protocol with  configurable reliability level as opposed to stacked two-layer \gls{arq}/\gls{harq} operation that is  currently deployed.

The rest of this paper is organized as follows: in \secref{Sec:Problem} the unreliability problem of retransmission protocols caused by    feedback channel unreliability feedback error problem is studied; \secref{Sec:Solution} introduces the  backwards composite feedback solution  for reliable packet delivery; in \secref{Sec:Results} numerical results are presented; to evaluate the performance of the proposed solution;
 finally, \secref{Sec:Conclusions} covers the concluding remarks.


\section{Problem description}
\label{Sec:Problem}

In this study we adopt the  \gls{saw}  mode of operation for retransmission protocols which works as follows. First, at time $i$ the $j$th packet arrived from a higher layer application denoted by $\pkt{i}{j}$ is transmitted by  	transmitter node for the first time. Next,  receiver node attempts decoding on the observed packet denoted by $\rpkt{i}{j}$. Using a feedback channel,  receiver node sends the  decoding success report $\tA{i}{}$ at corresponding feedback instance $i$, where  $\tA{i}{} = 1$ in case of \gls{ack} and $\tA{i}{} = 0$ in case of \gls{nack} (respectively, decoding success and decoding failure). Feedback transmission, similar to the data transmission, is assumed to be subject to channel impairments. We use $\rA{i}{}$ to denote the feedback observed by the transmitter node at feedback time instance $i$. In case of observing a \gls{nack} the same data packet \footnote{In practice the same message can be conveyed in different set of coded bits called \emph{redundancy versions}.} is retransmitted at the next transmit time instance $i+1$ (i.e., $\pkt{i+1}{j}$ ), otherwise, transmission of a new data packet is initiated (i.e., $\pkt{i+1}{j+1}$ ). Retransmission of a \gls{nack}ed packet continues until  \gls{ack} is observed over the feedback channel or maximum $M$ transmission attempts for the packet is reached. Therefore, at  transmitter node a packet is only regarded as \emph{delivered} if  \gls{ack} is observed and otherwise it is regarded as \emph{failed}. The  transmitter node sends a single-bit \gls{ndi} message per transmit data packet $\pkt{}{}$. The single-bit  \gls{ndi} is toggled every time a packet is transmitted for the first time. We assume that receiver is able to detect \gls{ndi}  error-free. The duration  between transmit occasions  $i$ and $i+1$ is denoted by \gls{rtt} where only one packet transmit occasion and one feedback occasion are considered in each \gls{rtt}.

Reliability of packet delivery in \gls{saw} operation  with  feedback channel is limited by both   packet transmission \gls{blep} and  feedback detection error rate. We use $\pe{} = \pe{1}$ and $\pe{m}$ for integer $m$ to denote  \gls{blep} of a packet after one and $m$ transmission attempts respectively where, by definition  $\pe{0} = 1$. We assume \gls{iid} block fading channel model for packet transmission. Therefore, we have $\pe{m} \leq (\pe{})^m$ where equality holds only if the decoder utilizes no combining gain (e.g., in case of \gls{arq} operation). Feedback channel is assumed to follow the \gls{bac} model where  error probability varies depending on the input symbol to the channel. Error probabilities for such channel model   are described as follows.
\begin{align}
\pn = \PR{\rA{i}{} = 1 | \tA{i}{} = 0}\\
\pa = \PR{\rA{i}{} = 0 | \tA{i}{} = 1}
\end{align}
We assume that instances of the feedback channel are independent from each other and from  data channel. Such model for the feedback channel is simplified as compared to real-life feedback channel where an extra message (e.g., \gls{dtx}) may also be considered as input to the feedback channel. E.g., in case of \gls{lte} technology, \gls{dtx} may  indicate failure in detection of the scheduling grant for data transmission \cite{3gpp36212}. Throughout this paper we reserve the notation $\bar{a}$ to denote $\bar{a} = 1 - a$ for any real valued $a$ where $a \in [0,1]$.

\tikzstyle{pkt}=[rectangle,
                                    thin,
                                    minimum width = 0.75cm,
                      				minimum height = 0.75cm,
                                    draw=black!80,
                                    fill=blue!25]
\tikzstyle{Empkt}=[rectangle,
                                    thin,
                                    minimum width = 0.75cm,
                      				minimum height = 0.75cm,
                                    draw=gray!30,
                                    fill=white!100]
\tikzstyle{DL}=[rectangle,
                                    thin,
                                    minimum width = 0.75cm,
                      				minimum height = 0.75cm,
                                    draw=gray!0,
                                    fill=white!0]
\tikzstyle{ret}=[rectangle,
                                    thin,
                                    minimum width = 0.75cm,
                      				minimum height = 0.75cm,
                      				draw=black!80,
                                    fill=gray!30]
\tikzstyle{ack}=[rectangle,
                                    thin,
                                    minimum width = 0.75cm,
                      				minimum height = 0.75cm,
                      				draw=black!80,
                                    fill=green!40,
                                    font=\tiny ,]
\tikzstyle{UL}=[rectangle,
                                    thin,
                                    minimum width = 0.75cm,
                      				minimum height = 0.75cm,
                      				draw=gray!0,
                                    fill=green!0,
                                    font=\tiny ,]
\tikzstyle{nack}=[rectangle,
                                    thin,
                                    minimum width = 0.75cm,
                      				minimum height = 0.75cm,
                                    draw=black!80,
                                    fill=red!40,
                                    font=\tiny ,]
\tikzstyle{boks}=[rectangle,
                                    thin,
                                    minimum width = 0.75cm,
                      				minimum height = 0.75cm,
                                    draw=gray!30,
                                    fill=white!100,
                                    font=\tiny ,]
\tikzstyle{Boks}=[rectangle,
                                    thin,
                                    minimum width = 0.75cm,
                      				minimum height = 0.75cm,
                                    draw=black!60,
                                    fill=white!100,
                                    font=\tiny ,]
\tikzstyle{background}=[rectangle,
                                    fill=gray!0,
                                    inner sep=0.2cm,
                                	rounded corners=5mm]


In a retransmission protocol where retransmissions are triggered by \gls{nack} feedback,  in case of \gls{nack}$\rightarrow$\gls{ack} error the transmitter node will mistakenly drop the packet assuming  it is successfully decoded at the receiver. Therefore, it is crucial to reduce the effective chances of a packet being discarded as a result of false \gls{ack}.  The straightforward solution to reduce chances of \gls{nack}$\rightarrow$\gls{ack} is to increase reliability of the feedback channel (i.e., reducing $\pn$) e.g., by increasing repetition order of feedback transmission by factor of $L > 1$. However, such solution  stretches  feedback message in time, frequency or power domains, requiring extra resources. Specifically, in scenarios where  receiver node has limited power and bandwidth for feedback transmission (e.g., narrow-band and low-cost  \gls{mmtc} receivers) the cost of increasing feedback reliability is  additional time diversity for  feedback which in turn  increases the experienced delay and receiver node power consumption. We use $T$ to denote the number of feedback occasions utilized for a packet before it is dropped at the transmitter (either considered \emph{delivered} or \emph{failed}). The average number of feedback occasions utilized per packet is then denoted by $\Tbar =  \mathop{\mathbb{E}} \{ T\}$ which, in time diversity scenario for feedback, is equivalent to the average delay experienced by the higher-layer application per packet. Further, we denote the events of decoding failure and success for $\pkt{}{j}$ in maximum $m$ transmission attempts by $\ef{j}{m}$ and $\es{j}{m}$ respectively. Outage probability, $\Pout$, is defined as the probability of decoding failure after maximum $M$ attempts, i.e., $\Pout =  \PR{\ef{j}{M}}$. Data channel utilization is measured by the average number  of transmission attempts per packet and is denoted by $\Nbar$. We use $\tau$ to denote the packet delivery latency defined as the time duration it takes from the first transmission attempt of a packet until it is correctly decoded at the receiver. Assuming zero processing time at the receiver node we have $\tau = \tr{TTI} + k \times \tr{RTT}$ where $k > 0$.

In the following we first study the effects of  feedback error on $\Pout$ performance of  retransmission protocols. Later in the next section we proposes a new feedback reporting scheme that suits low-cost and low-energy narrow-band wireless devices in ultra-reliable  packet delivery. We start by investigating  different feedback reporting approaches and analyze the trade-off between reliability and feedback time diversity order $L$.

\subsubsection{\gls{regsaw}}  
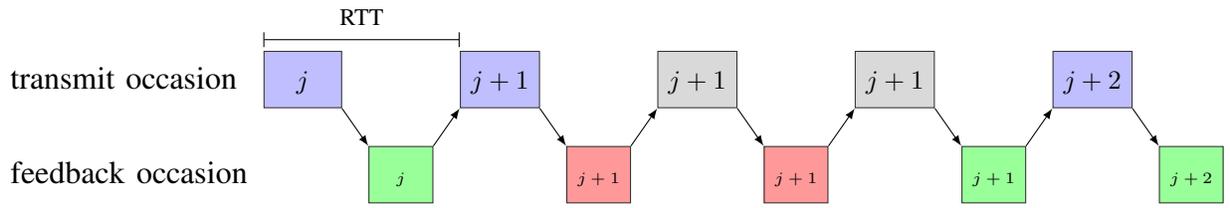
\begin{figure}[htbp]
\centering
\begin{tikzpicture}[>=latex,text height=1.5ex,text depth=0.25ex]
  \matrix[row sep=0.5cm,column sep=0.35cm] {
    \node	(p1)	[pkt]	{\small{$\; \; \; j \; \; \;$}}; &
    \node	(u1)	[UL]	{}; &
    \node	(p2)	[pkt]	{\small{$j+1$}}; &
    \node	(u2)	[UL]	{}; &
    \node	(p3)	[ret]	{\small{$j+1$}}; &
    \node	(u3)	[UL]	{}; &
    \node	(p4)	[ret]	{\small{$j+1$}}; &
    \node	(u4)	[UL]	{}; &
    \node	(p5)	[pkt]	{\small{$j+2$}}; &
    \\
    \node	(d1)	[DL]	{}; &
    \node	(f1)	[ack]	{$\; \; \; j \; \; \;$}; &
    \node	(d2)	[DL]	{}; &
    \node	(f2)	[nack]	{$j+1$}; &
    \node	(d3)	[DL]	{}; &
    \node	(f3)	[nack]	{$j+1$}; &
    \node	(d4)	[DL]	{}; &
    \node	(f4)	[ack]	{$j+1$}; &
    \node	(d5)	[DL]	{}; &
    \node	(f5)	[ack]	{$j+2$}; &
    \\
    };
\draw [->] (p1.south east) -- (f1.north west);
\draw [->] (p2.south east) -- (f2.north west);
\draw [->] (p3.south east) -- (f3.north west);
\draw [->] (p4.south east) -- (f4.north west);
\draw [->] (p5.south east) -- (f5.north west);
\draw [->] (f1.north east) -- (p2.south west);
\draw [->] (f2.north east) -- (p3.south west);
\draw [->] (f3.north east) -- (p4.south west);
\draw [->] (f4.north east) -- (p5.south west);
\draw[|-|] ($(p1.north west) + (0,0.15)$) to node[above] {$\ttr{RTT}$}($(p2.north west) + (0,0.15)$);
    \begin{pgfonlayer}{background}
        \node [background,
                    fit=(p1) (p4),
                    label=left:transmit occasion] {};
        \node [background,
                    fit=(d1) (f2),
                    label=left:feedback occasion] {};
    \end{pgfonlayer}
\end{tikzpicture}
\caption{Stop-and-wait operation.}
\label{Fig:SAWARQ}
\end{figure}

We assume the  \gls{bsc} model with $\pn = \pa = p$ for  feedback channel. As shown in \figref{Fig:SAWARQ}, in a \gls{saw} process, packet $\pkt{i}{j}$ is initially transmitted over $i$th (blue color)  transmit occasion which has duration of a \gls{tti}. The numbers inside the transmit occasion blocks indicates data packet index $j$ from $\pkt{i}{j}$ where $j$ is a positive integer.  Followed by transmission of the packet, after a given propagation  and receiver processing time \cite{khosravirad2016overview}, acknowledgement for the packet arrives.  Next,  transmitter node transmits the next packet $\pkt{i+1}{j+1}$ in case of \gls{ack} (green color blocks)  or retransmits a version of the same packet $\pkt{i+1}{j}$ (grey color  blocks) in case of \gls{nack} (red color blocks). \gls{ndi} is transmitted with each packet transmit occasion to notify the receiver of whether a new packet is transmitted (toggled \gls{ndi}) or the same packet is retransmitted (un-toggled \gls{ndi}). In principle  \gls{ack}  observance can be result of either a successful packet decoding followed by correct feedback detection or decoding failure followed by false feedback. In order to make it simple to follow the illustrations the packet index corresponding to each feedback occasion is shown inside the feedback occasion blocks.

The duration  between transmit occasions  $i$ and $i+1$ is denoted by \gls{rtt}. Without loss of generality, the propagation and processing time duration will be  skipped in the illustrations after \figref{Fig:SAWARQ}. Therefore, acknowledgement of each packet will be shown  below it while the next transmit occasion starts immediately after.

%
%

\subsubsection{\gls{Lrep}} To increase  feedback reliability for a receiver node with narrow-band low-energy feedback transmission a simple solutions is to increase time diversity of  feedback transmission by $L > 1$. In this model each feedback transmission  is stretched over $L$ feedback occasions where packet is declared as \emph{delivered} at the transmitter node only if all $L$ observances of feedback are  \gls{ack}. Otherwise,  the packet is retransmitted by the \gls{Lrep} process. Therefore, probability of false \gls{ack} reduces to $\pn^{L}$ compared to that of $\pn$ in case of \gls{regsaw}. Further, due to feedback repetition,  \gls{rtt} of \gls{Lrep} is $L$ times that of \gls{regsaw}. \gls{ndi} is used in similar way as in \gls{regsaw} to notify receiver node about retransmissions.

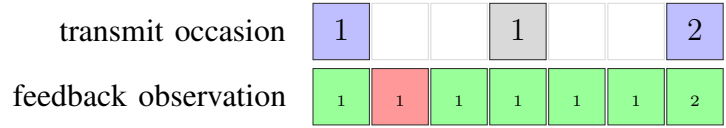
\begin{figure}[htbp]
\centering
\begin{tikzpicture}[>=latex,text height=1.5ex,text depth=0.25ex]
  \matrix[row sep=0.1cm,column sep=0.02cm] {
    \node	(p1)	[pkt]	{$1$}; &
    \node	(p2)	[Empkt]	{}; &
    \node	(p3)	[Empkt]	{}; &
    \node	(p4)	[ret]	{$1$}; &
    \node	(p5)	[Empkt]	{}; &
    \node	(p6)	[Empkt]	{}; &
    \node	(p7)	[pkt]	{$2$}; &
    \\
    \node	(f1)	[ack]	{$1$}; &
    \node	(f2)	[nack]	{$1$}; &
    \node	(f3)	[ack]	{$1$}; &
    \node	(f4)	[ack]	{$1$}; &
    \node	(f5)	[ack]	{$1$}; &
    \node	(f6)	[ack]	{$1$}; &
    \node	(f7)	[ack]	{$2$}; &
    \\
    };
    \begin{pgfonlayer}{background}
        \node [background,
                    fit=(p1) (p4),
                    label=left:transmit occasion] {};
        \node [background,
                    fit=(f1) (f2),
                    label=left:feedback observation] {};
    \end{pgfonlayer}
\end{tikzpicture}
\caption{\gls{Lrep} operation for $L = 3$.}
\label{Fig:LREPSAW}
\end{figure}

\subsubsection{\gls{Lack}} In this approach, the acknowledgment generated for a packet  is repeated over feedback occasions by the receiver node until $L > 1$ number of \gls{ack} observances are made at the transmitter node which in turn will trigger initiating the transmission of a new packet. A retransmission of the same packet is followed immediately if   \gls{nack} is observed while using \gls{ndi} receiver is notified about the retransmission. Note that in this approach  transmitter node keeps counting the number of \gls{ack} observances for a packet and  $L$ required \gls{ack} observances may be received in non-consecutive occasions  unlike  \gls{Lrep} approach where $L$  observances of \gls{ack} must be counted consecutively.

\begin{figure}[htbp]
\centering
\begin{tikzpicture}[>=latex,text height=1.5ex,text depth=0.25ex]
  \matrix[row sep=0.1cm,column sep=0.02cm] {
    \node	(p1)	[pkt]	{$1$}; &
    \node	(p2)	[Empkt]	{}; &
    \node	(p3)	[ret]	{$1$}; &
    \node	(p4)	[Empkt]	{}; &
    \node	(p5)	[pkt]	{$2$}; &
    \\
    \node	(f1)	[ack]	{$1$}; &
    \node	(f2)	[nack]	{$1$}; &
    \node	(f3)	[ack]	{$1$}; &
    \node	(f4)	[ack]	{$1$}; &
    \node	(f5)	[ack]	{$2$}; &
    \\
    };
    \begin{pgfonlayer}{background}
        \node [background,
                    fit=(p1) (p4),
                    label=left:transmit occasion] {};
        \node [background,
                    fit=(f1) (f2),
                    label=left:feedback observation] {};
    \end{pgfonlayer}
\end{tikzpicture}
\caption{\gls{Lack} operation for $L = 3$.}
\label{Fig:LACKSAW}
\end{figure}
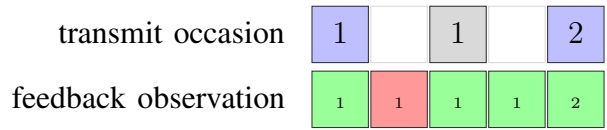

\subsubsection{\gls{RetxL}}
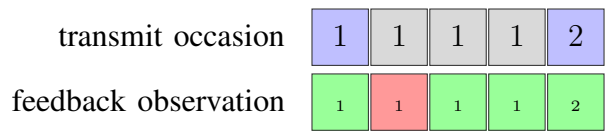
\begin{figure}[htbp]
\centering
\begin{tikzpicture}[>=latex,text height=1.5ex,text depth=0.25ex]
  \matrix[row sep=0.1cm,column sep=0.02cm] {
    \node	(p1)	[pkt]	{$1$}; &
    \node	(p2)	[ret]	{$1$}; &
    \node	(p3)	[ret]	{$1$}; &
    \node	(p4)	[ret]	{$1$}; &
    \node	(p5)	[pkt]	{$2$}; &
    \\
    \node	(f1)	[ack]	{$1$}; &
    \node	(f2)	[nack]	{$1$}; &
    \node	(f3)	[ack]	{$1$}; &
    \node	(f4)	[ack]	{$1$}; &
    \node	(f5)	[ack]	{$2$}; &
    \\
    };
    \begin{pgfonlayer}{background}
        \node [background,
                    fit=(p1) (p4),
                    label=left:transmit occasion] {};
        \node [background,
                    fit=(f1) (f2),
                    label=left:feedback observation] {};
    \end{pgfonlayer}
\end{tikzpicture}
\caption{\gls{RetxL} operation for $L = 3$.}
\label{Fig:ReTxLACK}
\end{figure}

In this approach, similar to \gls{Lack}, transmitter node requires $L$ observance of \gls{ack} before considering a packet as \emph{delivered}. However, transmitter continues retransmission of the packet while observing the feedback channel. Therefore,  \gls{RetxL} transmits each packet at least $L$ times and stops retranmission when $L$ times \gls{ack} observances are made or maximum $M$ transmission attempts is reached. \figref{Fig:ReTxLACK} depicts \gls{RetxL} process for $L = 3$ where retransmission of packet $\pkt{}{1}$ is continued until $3$ non-consecutive \gls{ack}s are detected.

\subsubsection{\gls{Asym}} A different approach  to decrease the false \gls{ack} rate is by  using asymmetric detection of the binary feedback channel. For instance, let's assume an \gls{awgn} feedback channel and a \gls{bpsk} symbol that conveys the single-bit feedback acknowledgement, where the binary $0$ and $1$ inputs of the feedback channel  are represented  in the signal constellation for \gls{bpsk} in terms of energy per bit $E_b$ respectively by $-\sqrt{E_b}$ and $\sqrt{E_b}$. Assuming coherent detection and perfect recovery of the carrier frequency and phase, from signal modulation and detection theory we know that the \gls{bep}  with symmetric decision regions \cite{Goldsmith:2005} is as follows, 
\begin{equation}
p =  \frac{1}{2}\erfc\bigg(\sqrt{\frac{E_b}{N_0}}\bigg),
\label{Eq:p}
\end{equation}
where $p$ denotes detection error probability, $N_0$ denotes the additive noise power spectral density and the complementary error function $\erfc{(\cdot)}$ is defined as, $ \erfc{(x)} = \frac{2}{\pi}\int_x^{\infty} \exp(-t^2) \, dt$.
Asymmetric decision regions, e.g., by moving the detection threshold in the \gls{bpsk} constellation from origin to the point $\alpha \times \sqrt{E_b}$  (closer to $\sqrt{E_b}$ than $-\sqrt{E_b}$ for $\alpha > 0$), decreases the modified chances of false \gls{ack} detection $\qn$, while the modified false \gls{nack} rate $\qa$  increases accordingly. This  reduces the chances of discarding unsuccessful packets at the transmitter while in turn increases chances of unnecessary retransmissions. The modified error probabilities for such asymmetric detection is then  as follows.
\begin{equation}
\qn = \frac{1}{2}\erfc{\bigg((1+\alpha)\sqrt{\frac{E_b}{N_0}}\bigg)}
\label{Eq:p0}
\end{equation}
\begin{equation}
\qa = \frac{1}{2}\erfc{\bigg((1-\alpha)\sqrt{\frac{E_b}{N_0}}\bigg)}
\label{Eq:p1}
\end{equation}
\gls{Asym} follows similar algorithm as in \gls{regsaw} where \gls{ndi} signal is utilized to notify receiver node of retransmissions. For performance evaluation of this approach we assume \gls{bpsk} modulation is used for the feedback channel where  $E_b$   is chosen based on a given $p$ in \eqref{Eq:p}. Then, the detection threshold is adjusted using parameter $\alpha$ in \eqref{Eq:p0} and \eqref{Eq:p1} to provide the required $\qn$.

\subsubsection{\gls{bretx}} We further investigate the performance of blind retransmission  without feedback. In such approach each packet is transmitted $M$ times by the transmitter node without requiring a feedback message from the receiver node.

Closed-form formulation for $\Pout$, $\Nbar$, $\Tbar$ and \gls{cdf} of packet delivery latency are shown in \tabref{Tab:FormulasI}, \tabref{Tab:FormulasII}, \tabref{Tab:FormulasIII} and \tabref{Tab:FormulasIV} respectively for the approaches described in this section. 

\def\arraystretch{2}
\begin{table*}[]
\centering
\caption{Outage probability, $\Pout$ for different feedback approaches. 
}
\label{Tab:FormulasI}
\begin{tabular}{l| c | c |}
\cline{2-3}
                       & finite $M$ & $M \rightarrow \infty $ \\ \hline
\multicolumn{1}{|l|}{\gls{regsaw}} & \begin{tabular}[l]{@{}l@{}} $\sum\limits_{m = 1}^{M-1} \pe{m} \pn \pnb^{m-1} + \pe{M} \pnb^{M-1}$ 
\end{tabular}  & \begin{tabular}[l]{@{}l@{}} 
$ \leq \pe{} \pn \frac{1}{1- \pe{} \pnb}$ \end{tabular}  \\ \hline
\multicolumn{1}{|l|}{$L$-Rep-ACK} & \begin{tabular}[l]{@{}l@{}} $\sum\limits_{m = 1}^{M-1} \pe{m} \pn^{L} (1-\pn^{L})^{m-1} + \pe{M} (1-\pn^{L})^{M-1}$ 
\end{tabular} & \begin{tabular}[l]{@{}l@{}} 
$\leq \pe{} \pn^{L} \frac{1}{1- \pe{} (1-\pn^{L})}$ \end{tabular} \\ \hline
\multicolumn{1}{|l|}{\gls{Lack}} & \begin{tabular}[l]{@{}l@{}}$\sum_{m=1}^{M-1}\pe{m}  \pn^{L} \pnb^{m-1} \binom{L+m-2}{m-1} +$ \\ $\pe{M} \pnb^{M-1} \sum_{l=0}^{L-1} \pn^{l} \binom{l+M-2}{M-2}$ \end{tabular}  & 
$\approx \pe{} \pn^{L} \big( 1 +   \frac{\pe{}\pnb}{(L-1)! \, (1-\pe{}\pnb)^L } \big)$ \\  \hline
\multicolumn{1}{|l|}{\gls{RetxL}} & \begin{tabular}[l]{@{}l@{}}$\sum_{m=L}^{M-1}\pe{m}  \pn^{L} \pnb^{m-L} \binom{m-1}{L-1} +$ \\ $\pe{M}  \pnb^{M-1} \sum_{l=0}^{L-1} (\frac{\pn}{\pnb})^{l}  \binom{M-1}{l}$ \end{tabular}  & 
$\approx \pe{L} \pn^{L} \big( 1 +   \frac{\pe{}\pnb}{(L-1)! \, (1-\pe{}\pnb)^L } \big)$ \\  \hline\multicolumn{1}{|l|}{\gls{Asym}} & \begin{tabular}[l]{@{}l@{}} $\sum\limits_{m = 1}^{M-1} \pe{m} \qn \qnb^{m-1} + \pe{M} \qnb^{M-1}$ 
\end{tabular} & \begin{tabular}[l]{@{}l@{}} 
$\leq \pe{} \qn \frac{1}{1- \pe{} \qnb}$ \end{tabular} \\ \hline
\multicolumn{1}{|l|}{Blind Retx} & $\pe{M}$ & $\rightarrow 0$  \\ \hline
\end{tabular}
\end{table*}


\def\arraystretch{2}
\begin{table*}[]
\centering
\caption{Average number of transmission attempt per packet, $\Nbar$}
\label{Tab:FormulasII}
\begin{tabular}{l| c | c |}
\cline{2-3}
                       & finite $M$ & $M \rightarrow \infty $ \\ \hline
\multicolumn{1}{|l|}{\gls{regsaw}} & \begin{tabular}[l]{@{}l@{}} $\sum\limits_{m = 1}^{M} \pe{m-1} \pnb^{m-1} +$ \\ $\sum\limits_{m=1}^{M-1} (\pe{m-1} - \pe{m}) \pnb^{m-1} \pa \frac{1-\pa^{M-m}}{\pab} $  \end{tabular}
& $ <  \frac{\pab + \bar{\pe{}} \pa }{\pab (1-\pe{}\pnb)} $  \\ \hline
\multicolumn{1}{|l|}{\gls{Lrep}} & \begin{tabular}[l]{@{}l@{}} $\sum\limits_{m = 1}^{M} \pe{m-1} (1 - \pn^L)^{m-1} +$ \\ $ \sum\limits_{m=1}^{M-1} (\pe{m-1} - \pe{m}) (1-\pn^L)^{m-1} (1-\pab^L) \frac{1-(1 - \pab^L)^{M-m}}{\pab^L} $  \end{tabular} & $ <  \frac{\pab^L + \bar{\pe{}} (1-\pab^L) }{ \pab^L (1-\pe{}(1-\pn^L))} $ \\ \hline
\multicolumn{1}{|l|}{\gls{Lack}} & \begin{tabular}[l]{@{}l@{}} $1+ g_M X_{L}^{M-1} + \bar{g_M} Y_{L}^{M-1} $\\ where, $\forall l \in \{1,...,L\}$ and $\forall m \in \{1,...,M-1\}$, \\ ${X_{l}^m} = \pnb \sum\limits_{\acute{l} = 1}^{l} \pn^{l-\acute{l}} \big( 1+ g_m X_{\acute{l}}^{m-1} + \bar{g_m} Y_{\acute{l}}^{m-1} \big) $,\\ $Y_l^m = \pa \sum\limits_{\acute{l}=1}^{l} \pab^{l-\acute{l}} (1+ Y_{\acute{l}}^m) $ \\ and, $X_l^0, Y_l^0 = 0$, $X_0^m, Y_0^m = 0$, and $g_m = \frac{\pe{M-m+1}}{\pe{M-m}}$ \end{tabular}  &  $< 1 +  \frac{\pe{} (1-\pnb^L) + \bar{\pe{}} (1-\pab^L)  }{\pab^L} $  \\  \hline
\multicolumn{1}{|l|}{\gls{RetxL}} & \begin{tabular}[l]{@{}l@{}} $\sum\limits_{m = L}^{M} m \eta_m$, where, \\$\eta_m = \pe{m} \pn^L \pnb^{m-L} \binom{m-1}{L-1} + \sum\limits_{l = 1}^{m} (\pe{l-1} - \pe{l}) \varrho_{m,l}$, with \\
$\varrho_{m,l} = \sum\limits_{k = \max\{L-m+l-1,0\}}^{\min\{l-1,L-1 \}} \frac{ \pn^k \pnb^{l} \pab^{L} \pa^{m+k+1}}{ \pnb^{k+1} \pab^{k} \pa^{L+l}} \binom{l-1}{k} \binom{m-l}{L-k-1}$ \\ and, \\ $\varrho_{M,l} = \sum\limits_{n = 0}^{L-1} \sum\limits_{k = \max\{n-M+l,0\}}^{\min\{l-1,n\}} \frac{ \pn^k \pnb^{l} \pab^{n} \pa^{M+k}}{ \pnb^{k+1} \pab^{k} \pa^{l+n}} \binom{l-1}{k} \binom{M-l}{n-k}$
 \end{tabular}  & ... \\  \hline
\multicolumn{1}{|l|}{\gls{Asym}} &  \begin{tabular}[l]{@{}l@{}} $\sum\limits_{m = 1}^{M} \pe{m-1} \qnb^{m-1} +$\\$ \sum\limits_{m=1}^{M-1} (\pe{m-1} - \pe{m}) \qnb^{m-1} \qa \frac{1-\qa^{M-m}}{\qab} $\end{tabular}   &  $ <  \frac{\qab + \bar{\pe{}} \qa }{\qab (1-\pe{}\qnb)} $ \\ \hline
\multicolumn{1}{|l|}{Blind Retx } & $M$ & $\infty$ \\ \hline
\end{tabular}
\end{table*}

\def\arraystretch{2}
\begin{table*}[]
\centering
\caption{Average experienced delay, $\Tbar$, in number of \gls{rtt}.}
\label{Tab:FormulasIII}
\begin{tabular}{l| c |}
\cline{2-2}
                       & assuming one feedback occasion per \gls{rtt} \\ \hline
\multicolumn{1}{|l|}{\gls{regsaw}} & $\Nbar$ \\ \hline
\multicolumn{1}{|l|}{\gls{Lrep}} & $\Nbar * L$  \\ \hline
\multicolumn{1}{|l|}{\gls{Lack}} & \begin{tabular}[l]{@{}l@{}} $1+ g_M X_{L}^{M-1} + \bar{g_M} Y_{L}^{M-1} $\\ where, $\forall l \in \{1,...,L\}$ and $\forall m \in \{1,...,M-1\}$, \\ ${X_{l}^m} = \pnb \sum\limits_{\acute{l} = 1}^{l} \pn^{l-\acute{l}} \big(l-\acute{l}+1+ g_m X_{\acute{l}}^{m-1} + \bar{g_m} Y_{\acute{l}}^{m-1} \big) $,\\ $Y_l^m = \pa \sum\limits_{\acute{l}=1}^{l} \pab^{l-\acute{l}} (1+ Y_{\acute{l}}^m) $ \\ and, $X_l^0, Y_l^0 = 0$, $X_0^m, Y_0^m = 0$, and $g_m = \frac{\pe{M-m+1}}{\pe{M-m}}$ \end{tabular}    \\  \hline
\multicolumn{1}{|l|}{\gls{RetxL}} &  $\Nbar$  \\ \hline
\multicolumn{1}{|l|}{\gls{Asym}} &  $\Nbar$  \\ \hline
\multicolumn{1}{|l|}{Blind Retx } & $M$  \\ \hline
\end{tabular}
\end{table*}

\def\arraystretch{2}
\begin{table*}[]
\centering
\caption{\gls{cdf} of packet delivery latency $\tau$.}
\label{Tab:FormulasIV}
\begin{tabular}{l| c |}
\cline{2-2}
                       & $\PR{\tau \leq \tr{TTI} + k * \tr{RTT} }$ \\ \hline
\multicolumn{1}{|l|}{\gls{regsaw}} & \begin{tabular}[l]{@{}l@{}} $ \pnb^{k} (\pe{k} - \pe{k+1}), \quad \forall k \in \{0,...,M-1\}$  \end{tabular} \\ \hline
\multicolumn{1}{|l|}{\gls{Lrep}} & \begin{tabular}[l]{@{}l@{}} $ (1-\pn^L)^{n} (\pe{n} - \pe{n+1}), \quad n = k*L, \, \forall k \in \{0,...,M-1\}$  \end{tabular} \\ \hline
\multicolumn{1}{|l|}{\gls{Lack}} & \begin{tabular}[l]{@{}l@{}}  $\sum\limits_{m = \max \{1,k-L+1\} }^{\min \{k, M-1 \} } \pnb^{k} (\pe{k} - \pe{k+1}) \pn^{k-m} \binom{l-1}{m-1}, \quad \forall k \in \{0,...,M+L-2\} $  \end{tabular}  \\  \hline
\multicolumn{1}{|l|}{\gls{RetxL}} & \begin{tabular}[l]{@{}l@{}} $ (\pe{k} - \pe{k+1})$ for $k \in \{0,...,L-1\}$, and \\
 $ \sum\limits_{\max\{1,k-L+1\}}^{k} (\pe{k} - \pe{k+1}) \pn^{k-m} \pnb^{m} \binom{k}{m} $ for $k \in \{L,...,M-1\}$\end{tabular} \\ \hline
\multicolumn{1}{|l|}{\gls{Asym}} &  \begin{tabular}[l]{@{}l@{}} $ \qnb^{k} (\pe{k} - \pe{k+1}), \quad \forall k \in \{0,...,M-1\}$  \end{tabular} \\ \hline
\multicolumn{1}{|l|}{Blind Retx } & $\pe{k} - \pe{k+1}, \quad \forall k \in \{0,...,M-1\}$ \\ \hline
\end{tabular}
\end{table*}


\begin{figure}[t]
\begin{center}
\psfrag{xlabel}[c][c][\scalevalueS]{$p$}
\psfrag{ylabel}[c][c][\scalevalueS]{$\Pout$}
\psfrag{XXXXXXXXXXXXXXXXX1}[lc][lc][\scalevalueSS]{\gls{regsaw}}
\psfrag{x2}[lc][lc][\scalevalueSS]{$2$-Rep-ACK}
\psfrag{x3}[lc][lc][\scalevalueSS]{$4$-Rep-ACK}
\psfrag{x4}[lc][lc][\scalevalueSS]{$2$-ACK-SAW}
\psfrag{x5}[lc][lc][\scalevalueSS]{$4$-ACK-SAW}
\psfrag{x6}[lc][lc][\scalevalueSS]{\gls{Asym}, $\qn = \min \{10^{-3},p\}$}
\psfrag{x7}[lc][lc][\scalevalueSS]{\gls{Asym}, $\qn = \min \{10^{-4},p\}$}
\psfrag{XXXXXXXXX8}[lc][lc][\scalevalueSS]{ReTx-$2$-ACK}
\psfrag{x9}[lc][lc][\scalevalueSS]{ReTx-$4$-ACK}
\psfrag{r1}[lc][lc][\scalevalueS]{$R1$}
\psfrag{r2}[lc][lc][\scalevalueS]{$R2$}
\psfrag{r3}[lc][lc][\scalevalueS]{$R3$}
\psfrag{r4}[lc][lc][\scalevalueS]{$R4$}
\psfrag{r5}[lc][lc][\scalevalueS]{$R5$}
\psfrag{r6}[lc][lc][\scalevalueS]{$R6$}
\includegraphics[width=.9\columnwidth,keepaspectratio]{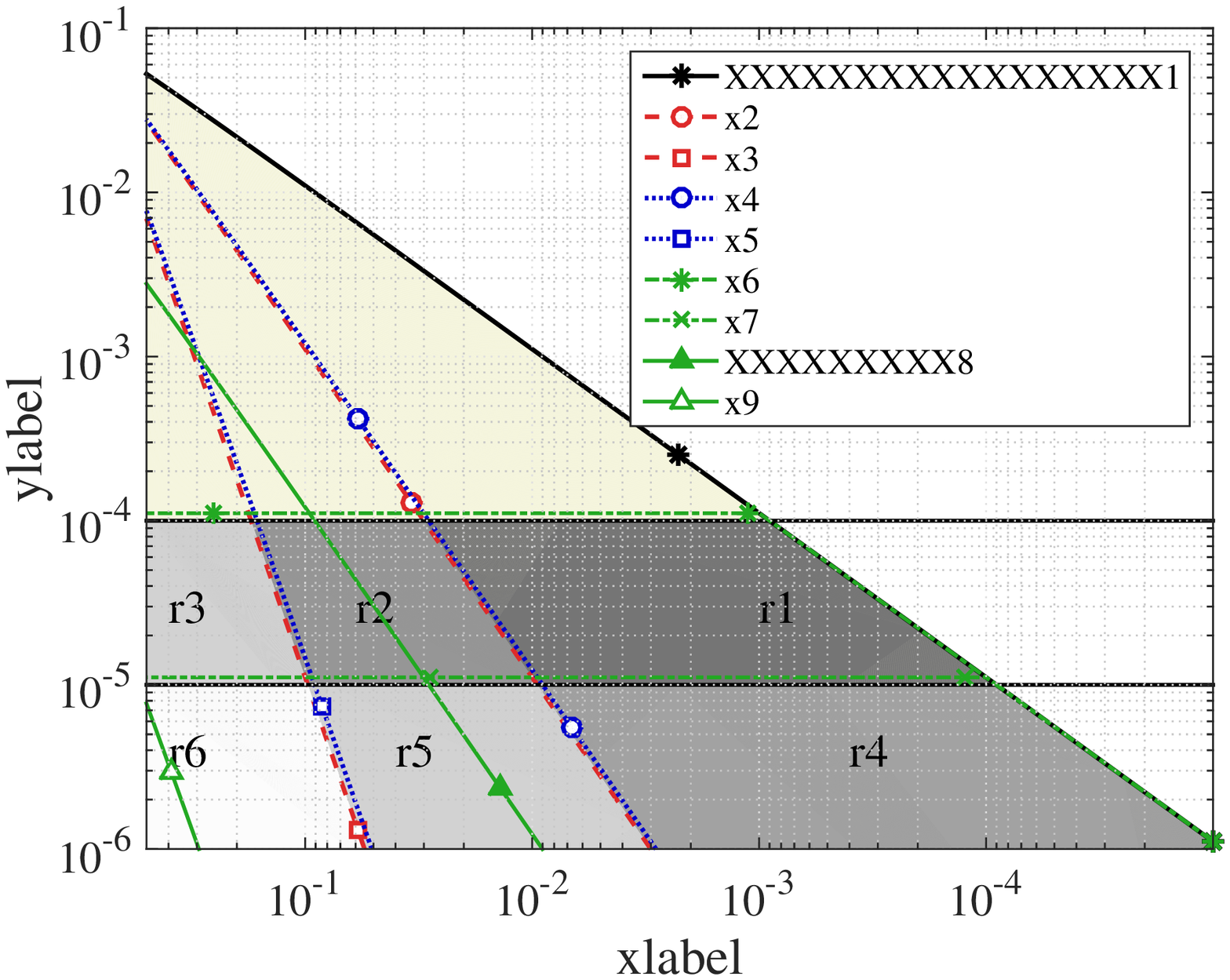}
\caption{Outage probability for $M \rightarrow \infty $ against a range of $p = \pn = \pa$ values with $\pe{} = 0.1$.}
\label{Fig:PoutRegions}
\end{center}
\vspace{-20pt}
\end{figure}


For an infinite allowed number of transmission attempts ($M \rightarrow \infty$) and assuming zero combining gain at the receiver (i.e., $\pe{m} =(\pe{})^m$), in \figref{Fig:PoutRegions} the outage probability $\Pout$ of the above-listed feedback approaches is illustrated. Reliability of  \gls{regsaw} is proportional to feedback reliability metric $p$ even in unbounded $M$ scenario. Specifically, the operation regions on \figref{Fig:PoutRegions} that are labeled by $R1, ..., R6$ for $\Pout$ below $10^{-4}$ and $10^{-5}$  are not achievable using \gls{regsaw}. Therefore, for ultra-reliable communication it is required to either increase  feedback diversity order $L$ or to perform blind retransmission without reliance on unreliable feedback channel. By increasing  feedback time diversity  order by $L = 2$,  reliability regions $R1$ and $R4$ are achievable with proper choice of $M$. However, achieving  regions $R2$ and $R5$ requires $L>2$. Interestingly, \gls{bretx} with $M = 5$ and $6$ can achieve reliability in regions $R2$ and $R5$ respectively by refusing the dependency on feedback channel. However, as we  see later in this paper such approach can be harmfully inefficient in resource utilization. Nevertheless, for highly unreliable feedback channel conditions such as in region $R6$, very large feedback diversity  order $L>>4$ can have reverse effect on the performance efficiency parameters such as channel utilization and average delay. This  makes \gls{bretx} a viable solution for when the feedback channel is unable to offer a reasonable level of reliability. Further,  \gls{Asym} feedback operation requires stringent $\qn$ adjustment of, e.g.,   $\qn < 10^{-4}$ for $\Pout < 10^{-5}$. In highly unreliable feedback channel conditions, with $\qn \rightarrow 0$,  false \gls{nack} rate increases drastically (i.e., $\qa \rightarrow 1$) which increases  number of transmission attempts resulting in similar performance as \gls{bretx} approach. 

Moreover, from \figref{Fig:PoutRegions} it is observed that  reliability performance of  \gls{Lrep} and \gls{Lack} approaches are tightly similar in practical range of feedback diversity order $L$. 
 The  downside of increasing feedback diversity order is the increased energy consumption at the receiver to report more than one feedback per packet transmission (i.e., $\Tbar$). 
  In particular, for  use cases where battery life-time is of critical importance, less  energy consumption over feedback reports is desirable. Thus, it is required to reduce feedback energy consumption   while configuring  high reliability for the  retransmission protocol.  This motivates  next section of this paper where we propose a variant of the \gls{Lack} approach which similarly requires $L$ observances of \gls{ack} for a data packet to be considered as \emph{delivered}. However, the new solution reduces  number of average feedback occasions used per packet thanks to the proposed backwards composite feedback operation. This way,  $\Tbar$ improves compared to \gls{Lack}  while thanks to the required multiple \gls{ack} observance per packet, a higher reliability of operation is expected compared to \gls{regsaw}.

%


\section{Backwards composite feedback}
\label{Sec:Solution}

In this section we propose a composite feedback solution to provide highly-reliable \gls{saw} operation in unreliable  feedback channel condition, which can be applied to  retransmission protocols such as \gls{arq}/\gls{harq}. In the proposed \gls{bcf} solution, the aim is to observe a given  $L > 1$ times  \gls{ack}  for a  packet before the packet is labeled reliably as \emph{delivered}.  In order to avoid drastic increase in $\Tbar$, in the proposed \gls{bcf} solution we suggest to repeat the feedback for each packet in a composite manner. We assume that a \gls{bcf} process has at most $L$ \emph{active packets} in its buffers at each time instance. An active packet is identified as a packet that has been transmitted $m$ times, where $1 \leq m < M$, and \gls{ack} feedback is observed for it less than $L$ times. 
We define  composite feedback at the receiver node at time $i$ as follows, where $l$ denotes  index of the active packets set.
\begin{align} \label{Eq:CompFeedback}
\tC{i} = \underset{l}{\&}\tA{}{l} = \left\{\begin{matrix}
1 & \tA{}{l} = 1, \; \forall l \\ 
0 & \mbox{otherwise}
\end{matrix}\right.
\end{align}
Therefore, an observed composite feedback $\rC{i} = 1$ at the transmitter is counted as \gls{ack} for all  active packets. In the case  $\rC{i} = 0$ is observed, a retransmission phase cycle starts which attempts on retransmitting active packets one after another following a given order of packets. The retransmission phase  then continues until $\rC{} = 1$ is observed or the maximum transmission attempt is reached for all  active packets. The propose \gls{bcfsaw} operation at  transmitter and receiver nodes is as follows.

\vspace{-5pt}
\begin{algorithm}
\begin{footnotesize}
    \SetKwInOut{Input}{Input}
    \SetKwInOut{Output}{Output}
    \Input{observed composite feedback $\rC{i}$}
    \Output{transmit packet $\pkt{i}{}$; new data indicator $\mbox{NDI}_i$}
  \uIf{$\rC{i} == 1$}{
    $\mbox{ACKcounter}(l)++ \quad \forall l$\;
    $\mbox{NDI}_{i+1} \leftarrow \mbox{\textbf{toggle}}(\mbox{NDI}_i)$\;
    $\mbox{NDItoggle}++ \mod L$\;
    $\mbox{Buffer}(\mbox{NDItoggle}) \leftarrow$\textbf{get new packet} \;
    $\mbox{ACKcounter}(\mbox{NDItoggle}) = 0$\;
    $\mbox{NAKcounter}(\mbox{NDItoggle}) = 0$\;
    $\pkt{i+1}{} \leftarrow \mbox{Buffer}(\mbox{NDItoggle})$\;
    $\mbox{TXcountre}(\mbox{NDItoggle}) = 1$\;
    \textbf{clear} Indx\;
  }
  \Else{
  	\uIf{$\mbox{{TXcounter}}(l) == 0 \; \mbox{or} \; M \quad \forall l$,}{
  	\textbf{go to}  3 	
  	}
  	\Else{
  	$\mbox{NDI}_{i+1} \leftarrow \mbox{NDI}_i$\;
    $\mbox{NAKcounter}(l)++ \quad \forall l$\;
    $\mbox{Indx} \leftarrow$ \textbf{look up reTx index}$\big( \mbox{TXcounter},$ $\mbox{ACKcounter},$ $\mbox{NAKcounter} \big)$\;
    $\pkt{i+1}{} \leftarrow \mbox{Buffer}(\mbox{Indx})$\;
    $\mbox{TXcountre}(\mbox{Indx})++$\;
    }
  }
    $i++$\;
    \textbf{return} $\pkt{i}{}$, $\mbox{NDI}_i$\;  
    \textbf{go to} 1\;
    \caption{Operation at the transmitter}
\label{Alg:Tx}
\end{footnotesize}
\end{algorithm}
\vspace{-10pt}

\subsection{Operation at the transmitter}

\Algref{Alg:Tx} presents the \gls{bcf} algorithm at the transmitter side. We assume that all the active packets are stored in separate buffers at the transmitter for in case a retransmission is needed. An active packet is then represented by $\mbox{Buffer}(l)$ for $l \in [0,...,L-1]$. The variable  NDItoggle stores the index $l$ of the last active packet which was transmitted for the first time (i.e., NDI was toggled for it).  When  transmitter node observes  \gls{ack} over the feedback channel it increments  counters $\mbox{ACKcounter}(l)$  by one for all $l$.  Then, NDI is toggled and  NDItoggle index is incremented by one $\mod L$. The updated NDItoggle points either to an empty buffer or to an active packet where  $\mbox{ACKcounter}(\mbox{NDItoggle}) = L$.  Buffer(NDItoggle) is therefore reset and substituted by a new packet taken from the higher layer application (this process is denoted by function \textbf{get new packet}). Next transmit occasion is then  utilized to transmit  the newly initiated active packet. A toggle in NDI bit informs the receiver node about  transmission of a new packet.

In the case where \gls{nack} is observed, the retransmission phase of the operation starts where  active packets are retransmitted one after another following a given order until \gls{ack} is observed or the maximum transmission attempt is reached for all active packets.  The order in which  active packets are retransmitted in this phase is given in a look up table that is pre-shared between  receiver and  transmitter nodes. The look-up table identifies the next active packet index denoted by variable Indx that is to be retransmitted. This process is denoted by function \textbf{look up reTx index} which inputs counters TXcounter($l$), ACKcounter($l$) and NAKcounter($l$) respectively denoting  number of transmission attempts, observed \gls{nack}s and observed \gls{ack}s for active packet $l$.  NDI signal  remains untoggled during retransmission phase.

\vspace{-5pt}
\begin{algorithm}
\begin{footnotesize}
    \SetKwInOut{Input}{Input}
    \SetKwInOut{Output}{Output}
    \Input{observed NDI; received packet $\rpkt{i}{}$}
    \Output{composite feedback $\tC{i}$}
  \uIf{NDI is toggled}{
    $\mbox{ACKcounter}(l)++ \quad \forall l \in [0,...,L-1]$\;
    $\mbox{NDItoggle}++ \mod L$\;
    $\mbox{Buffer}(\mbox{NDItoggle}) \leftarrow \rpkt{i}{}$\;
    $\mbox{RXcountre}(\mbox{NDItoggle}) = 1$\;
    $\tA{}{\mbox{NDItoggle}} \leftarrow \mbox{\textbf{decode success}}(\rpkt{i}{})$
  }
  \Else{
    $\mbox{NAKcounter}(l)++ \quad \forall l$\;
    $\mbox{Indx} \leftarrow$ \textbf{look up reTx index}$\big( \mbox{RXcounter},$ $\mbox{ACKcounter},$ $\mbox{NAKcounter} \big)$\;
    $\mbox{RXcountre}(\mbox{Indx})++$\;
    $\tA{}{\mbox{Indx}} \leftarrow \mbox{\textbf{decode success}}(\rpkt{i}{},\mbox{Buffer}(\mbox{Indx}))$\;
    $\mbox{Buffer}(\mbox{Indx}) \leftarrow \mbox{\textbf{combine}}(\rpkt{i}{},\mbox{Buffer}(\mbox{Indx}))$\;
  }
    $i++$\;    
    $\tC{i} \leftarrow \underset{l}{\&}\tA{}{l}$ where $l = \mbox{NDItoggle} \, \mbox{or,} \, \mbox{RXcounter}(l)\neq 0, M$ \;
    \textbf{return} $\tC{i}$\;
  \textbf{go to} 1\;
    \caption{Operation at the receiver}
\label{Alg:Rx}
\end{footnotesize}
\end{algorithm}
\vspace{-10pt}

\subsection{Operation at the receiver}

Receiver node follows \Algref{Alg:Rx} where firstly it detects whether the received packet is a new transmission or it is retransmission of one of the earlier received active packets. This is done at each time $i$ by observing $\mbox{NDI}_i$ signal and comparing it with $\mbox{NDI}_{i-1}$. The variable $\mbox{NDItoggle} \in [0,...,L-1]$ store the index of the active packet which is most recently received for the first time.  If NDI is detected to be toggled,  NDItoggle index at the receiver is incremented by one $\mod L$. Then, $\mbox{Buffer}(\mbox{NDItoggle})$ is substituted with the newly received packet. Function \textbf{decode success} outputs the feedback generated from decoding $l$th active packet, denoted using upper case index by $\tA{}{l}$. We assume zero chance of error detection failure where the acknowledgement  for a packet is generated based on error detection for the packet, e.g., using \gls{crc}. 

In case  NDI signal is detected as untoggled, the algorithm follows retransmission phase operation where  the pre-shared look-up table is used to find the index of the active packet that is being retransmitted (denoted by Indx). Decoding in such case may be based on the received retransmission and the stored version of the active packet from previous (re)transmission attempts to provide the decoder with combing gain. Similarly, the buffer content may be updated (e.g., in case  of $\tA{}{\tr{Indx}} = 0$) after combining the two versions of the packet. Composite feedback $\tC{i}$ is generated using  $\tA{}{l}$ for all the active packet indexes $l$ according to \eqref{Eq:CompFeedback}.

%
%

The retransmission phase order of packets  follows a pre-shared look-up table. As discussed above we assume that both nodes keep track of the number of transmission attempts and the number of observed \gls{ack} and \gls{nack} for each active packet. Given the value of these counters the next packet to be retransmitted in case of \gls{nack} is found.

\subsection{Example case for $L = 2$}

The proposed \gls{bcf} operation is explained below for the case of $L = 2$, i.e., twice \gls{ack} observances  required for a packet to be considered as \emph{delivered}). The cases of larger $L$ will follow a similar approach.

\subsubsection*{Operation in case of observed composite \gls{ack}, $\rC{i} = 1$}
As depicted in \figref{Fig:BFB-ACK}, two  active packets are assumed at the transmitter each having a separate TXcounter and  ACKcounter. The process starts by transmitting packet $\pkt{1}{1}$ thus, the first feedback occasion only acknowledges  decoding status for packet 1, $\tC{1} = \tA{}{1}$.  Transmitter then initiates the second active packet by transmitting packet $\pkt{2}{2}$ in the following transmit occasion. We assume that  receiver is notified of the new packet using a single-bit \gls{ndi} signal. 

In the second feedback occasion,  receiver composites  acknowledgement the two active packets resulting in composite \gls{nack} due to decoding failure for $\pkt{2}{2}$, $\tC{2} = \tA{}{1} \& \tA{}{2} = 0$. The observed composite feedback at time $i = 2$ is however erroneously detected as \gls{ack}, i.e., $\rC{2} = 0$. The observed \gls{ack} feedback counts as one \gls{ack} for both active packets resulting. As a result, ACKcounter for packet 1 reaches $L = 2$ and the packet is regarded as delivered and thus discarded from  \gls{bcfsaw}  process. Note that for ease of following the illustration, TXcounter blocks in \figref{Fig:BFB-ACK} show the counter value and the corresponding packet number in brackets.

Next, $\pkt{3}{3}$ is transmitted and decoded successfully. However, since $\tA{}{2} = 0$, the composite feedback at time $i = 3$ is $\tC{3} = 0$. A second feedback error at time $i = 3$ results in $\rC{3} = 1$ and as a result $\pkt{}{2}$ is discarded from the transmitter buffer even though it failed in decoding. Such outage case requires  $L = 2$ times \gls{nack}$\rightarrow$\gls{ack} errors during the time a packet is in an active packet buffer of the transmitter node.

\begin{figure}[htbp]
\centering
\begin{tikzpicture}[>=latex,text height=1.5ex,text depth=0.25ex]
\matrix[row sep=0.1cm,column sep=0.02cm] {
    \node	(p1)	[pkt]	{$1$}; &
    \node	(p2)	[pkt]	{$2$}; &
    \node	(p3)	[pkt]	{$3$}; &
    \node	(p4)	[pkt]	{$4$}; &
    \node	(p5)	[pkt]	{$5$}; &
    \\
    \node	(c1)	[ack]	{$1$}; &
    \node	(c2)	[nack]	{$1\&2$}; &
    \node	(c3)	[nack]	{$2\&3$}; &
    \node	(c4)	[ack]	{$3\&4$}; &
    \node	(c5)	[ack]	{$4\&5$}; &
    \\ 
    \node	(f1)	[ack]	{$1$}; &
    \node	(f2)	[ack]	{$1\&2$}; &
    \node	(f3)	[ack]	{$2\&3$}; &
    \node	(f4)	[ack]	{$3\&4$}; &
    \node	(f5)	[ack]	{$4\&5$}; &
    \\ 
    \\
    \node	(b01)	[Boks]	{$1 (1)$}; &
    \node	(b2)	[boks]	{$1 (1)$}; &
    \node	(b3)	[Boks]	{$1 (3)$}; &
    \node	(b4)	[boks]	{$1 (3)$}; &
    \node	(b5)	[Boks]	{$1 (5)$}; &
    \\
    \node	(b11)	[Boks]	{$1$}; &
    \node	(b2)	[boks]	{$2$}; &
    \node	(b3)	[Boks]	{$1$}; &
    \node	(b4)	[boks]	{$2$}; &
    \node	(b5)	[Boks]	{$1$}; &
    \\ \\
    \node	(b21)	[boks]	{$0$}; &
    \node	(b2)	[Boks]	{$1 (2)$}; &
    \node	(b3)	[boks]	{$1 (2)$}; &
    \node	(b4)	[Boks]	{$1 (4)$}; &
    \node	(b5)	[boks]	{$1 (4)$}; &
    \\
    \node	(b31)	[boks]	{$0$}; &
    \node	(b2)	[Boks]	{$1$}; &
    \node	(b3)	[boks]	{$2$}; &
    \node	(b4)	[Boks]	{$1$}; &
    \node	(b5)	[boks]	{$2$}; &
    \\
    };
    \begin{pgfonlayer}{background}
        \node [background,
                    fit=(p1) (p4),
                    label=left:transmit occasion] {};
        \node [background,
                    fit=(c1) (c2),
                    label=left:composite feedback] {};
        \node [background,
                    fit=(f1) (f2),
                    label=left:feedback observation] {};                
        \node [background,
                    fit=(b01),
                    label=left:transmitter TXcounter(1)] {};
        \node [background,
                    fit=(b11),
                    label=left:transmitter ACKcounter(1)] {};
        \node [background,
                    fit=(b21),
                    label=left:transmitter TXcounter(2)] {};
        \node [background,
                    fit=(b31),
                    label=left:transmitter ACKcounter(2)] {};
    \end{pgfonlayer}
\end{tikzpicture}
\caption{Backwards feedback bundling operation in case of composite \gls{ack}.}
\label{Fig:BFB-ACK}
\end{figure}

\subsubsection*{Operation in case of observed  composite \gls{nack}, $\rC{i} = 0$}

An observed composite \gls{nack} initiates retransmission phase where   active packets are retransmitted one after another according to a pre-shared order of packets. \gls{ndi}  remains un-toggled during  retransmission phase.  The retransmission phase continues  until  \gls{ack} is observed or the transmit counter for all active packets reaches $M$. In \figref{Fig:BFB-NACK} we assume similar events have encountered as in \figref{Fig:BFB-ACK} up to feedback occasion $i = 3$. Let's assume that at $i = 3$ in \figref{Fig:BFB-NACK} composite feedback is correctly detected, $\rC{3} = 0$. From transmitter point of view, observed composite \gls{nack} may be caused by several events including decoding failure of one of the active packets or feedback channel error.  For instance, the observed composite \gls{nack} $\rC{3}$ in \figref{Fig:BFB-NACK} may be the result of any of the following events.
\begin{itemize}
\item	$E1$: $F_3^1 \& S_2^1$  followed by successful  feedback detection at $i = 3$
\item	$E2$: $F_2^1 \& S_3^1$ with successful  feedback detection at $i = 3$ and feedback  error at $i = 2$
\item	$E3$: $F_2^1 \& F_3^1$ with successful  feedback detection at $i = 3$ and feedback  error at $i = 2$
\item	$E4$: $S_2^1 \& S_3^1$  followed by feedback  error at $i = 3$
\end{itemize}
The likelihood of such events can be evaluated as it was explained in \secref{Sec:RetPacketOrder}. E.g., for  packet transmission with target $\pe{} = 0.1$ and feedback channel reliability of $p = 0.001$, $E1$ is the more likely event making $\pkt{}{3}$ the best candidate for retransmission in next transmit occasion. Nevertheless, the retransmission packet order as a function of TXcounter, ACKcounter and NAKcounter can be established prior to start of the communication process and shared between communicating nodes.

In \figref{Fig:BFB-NACK}, retransmission of $\pkt{4}{3}$ is performed at $i = 4$. Due to failure of $\pkt{2}{2}$, \gls{nack} composite feedback $\tC{4} = 0$ is generated at the receiver node and correctly detected at the transmitter. Therefore, retransmission phase continues and transmitter uses the look-up table to pull the next index of active packet that must be retransmitted. Thus, at time $i = 5$, $\pkt{5}{2}$ is retransmitted resulting in $S_2^2$.

\begin{figure}[htbp]
\centering
\begin{tikzpicture}[>=latex,text height=1.5ex,text depth=0.25ex]
\matrix[row sep=0.1cm,column sep=0.02cm] {
    \node	(p1)	[pkt]	{$1$}; &
    \node	(p2)	[pkt]	{$2$}; &
    \node	(p3)	[pkt]	{$3$}; &
    \node	(p4)	[ret]	{$3$}; &
    \node	(p5)	[ret]	{$2$}; &
    \\
    \node	(c1)	[ack]	{$1$}; &
    \node	(c2)	[nack]	{$1\&2$}; &
    \node	(c3)	[nack]	{$2\&3$}; &
    \node	(c4)	[nack]	{$2\&3$}; &
    \node	(c5)	[ack]	{$2\&3$}; &
    \\
    \node	(f1)	[ack]	{$1$}; &
    \node	(f2)	[ack]	{$1\&2$}; &
    \node	(f3)	[nack]	{$2\&3$}; &
    \node	(f4)	[nack]	{$2\&3$}; &
    \node	(f5)	[ack]	{$2\&3$}; &
    \\ 
    \\
    \node	(b01)	[Boks]	{$1 (1)$}; &
    \node	(b2)	[boks]	{$1 (1)$}; &
    \node	(b3)	[Boks]	{$1 (3)$}; &
    \node	(b4)	[Boks]	{$2 (3)$}; &
    \node	(b5)	[boks]	{$2 (3)$}; &
    \\
    \node	(b11)	[Boks]	{$1$}; &
    \node	(b2)	[boks]	{$2$}; &
    \node	(b3)	[Boks]	{$0$}; &
    \node	(b4)	[Boks]	{$0$}; &
    \node	(b5)	[boks]	{$1$}; &
    \\ \\
    \node	(b21)	[boks]	{$0$}; &
    \node	(b2)	[Boks]	{$1 (2)$}; &
    \node	(b3)	[boks]	{$1 (2)$}; &
    \node	(b4)	[boks]	{$1 (2)$}; &
    \node	(b5)	[Boks]	{$2 (2)$}; &
    \\
    \node	(b31)	[boks]	{$0$}; &
    \node	(b2)	[Boks]	{$1$}; &
    \node	(b3)	[boks]	{$1$}; &
    \node	(b4)	[boks]	{$1$}; &
    \node	(b5)	[Boks]	{$2$}; &
    \\
    };
    \begin{pgfonlayer}{background}
        \node [background,
                    fit=(p1) (p4),
                    label=left:transmit occasion] {};
        \node [background,
                    fit=(c1) (c2),
                    label=left:composite feedback] {};
        \node [background,
                    fit=(f1) (f2),
                    label=left:feedback occasion] {};
        \node [background,
                    fit=(b01),
                    label=left:transmitter TXcounter(1)] {};
        \node [background,
                    fit=(b11),
                    label=left:transmitter ACKcounter(1)] {};
        \node [background,
                    fit=(b21),
                    label=left:transmitter TXcounter(2)] {};
        \node [background,
                    fit=(b31),
                    label=left:transmitter ACKcounter(2)] {};
    \end{pgfonlayer}
\end{tikzpicture}
\caption{Backwards composite feedback  operation in case of observed \gls{nack}. Feedback error has occurred in the second feedback occasion.}
\label{Fig:BFB-NACK}
\end{figure}

The proposed \gls{bcfsaw} uses the same number of feedback occasions per packet transmit occasions as in  \gls{regsaw} in \figref{Fig:SAWARQ} without the need to increase time diversity of feedback channel. However, each  packet is \gls{ack}ed $L$ times over the feedback channel which  in turn increases the reliability of packet delivery.

\begin{figure}[]
\begin{center}
\psfrag{xlabel}[c][c][\scalevalueS]{$p$}
\psfrag{ylabel}[c][c][\scalevalueS]{outage probability, $\Pout$}
\psfrag{XXXXXXXXX1}[lc][lc][\scalevalueSS]{\gls{bcf}-\gls{arq}, $L = 2$}
\psfrag{x2}[lc][lc][\scalevalueSS]{\gls{bcf}-\gls{arq}, $L = 4$}
\psfrag{x3}[lc][lc][\scalevalueSS]{\gls{bcf}-\gls{harq}, $L = 2$}
\psfrag{x4}[lc][lc][\scalevalueSS]{\gls{bcf}-\gls{harq}, $L = 4$}
\psfrag{x5}[lc][lc][\scalevalueSS]{$2$-ACK-ARQ}
\psfrag{x6}[lc][lc][\scalevalueSS]{$4$-ACK-ARQ}
\psfrag{x7}[lc][lc][\scalevalueSS]{$2$-ACK-HARQ}
\psfrag{XXXXXXXXX8}[lc][lc][\scalevalueSS]{$4$-ACK-HARQ}
\psfrag{x11}[lc][lc][\scalevalueSS]{Asym-\gls{harq}, $\qn = \min \{10^{-5},p\}$}
\psfrag{XXXXXXXXXXXXXXX12}[lc][lc][\scalevalueSS]{Asym-\gls{harq}, $\qn = \min \{10^{-4},p\}$}
\psfrag{r1}[lc][lc][\scalevalueS]{\color[RGB]{153, 50, 204}{$R1$}}
\psfrag{r2}[lc][lc][\scalevalueS]{\color[RGB]{153, 50, 204}{$R2$}}
\psfrag{r3}[lc][lc][\scalevalueS]{\color[RGB]{153, 50, 204}{$R3$}}
\psfrag{r4}[lc][lc][\scalevalueS]{\color[RGB]{153, 50, 204}{$R4$}}
\psfrag{r5}[lc][lc][\scalevalueS]{\color[RGB]{153, 50, 204}{$R5$}}
\psfrag{r6}[lc][lc][\scalevalueS]{\color[RGB]{153, 50, 204}{$R6$}}
\includegraphics[width=\columnwidth,keepaspectratio]{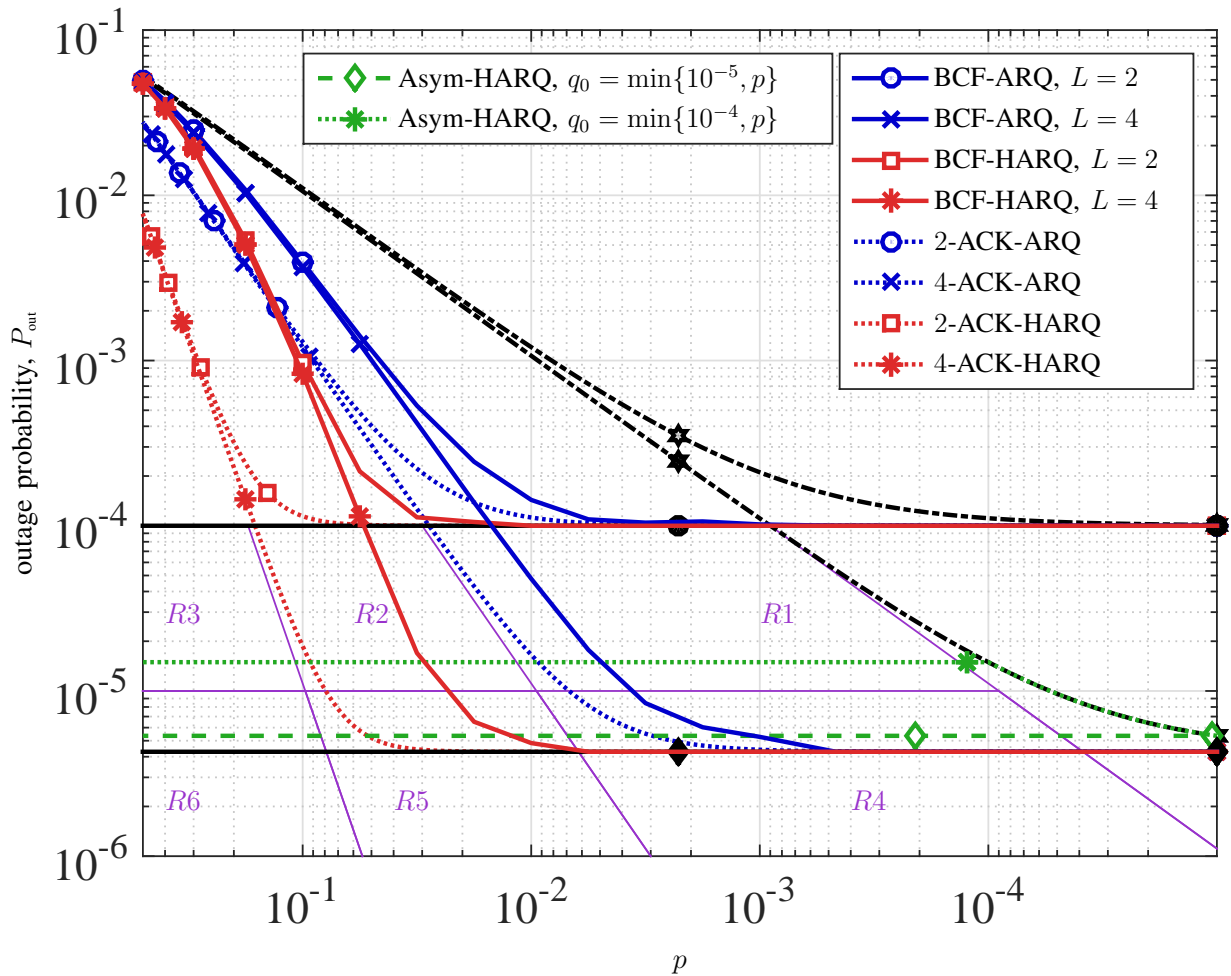}
\caption{Outage probability for $\pn = \pa = p$. The four black-colored markers from top to bottom represent  \gls{regsaw} for \gls{arq} and \gls{harq} and \gls{bretx} for \gls{arq} and \gls{harq}.}
\label{Fig:Outage}
\end{center}
\vspace{-20pt}
\end{figure}

\section{Numerical results}
\label{Sec:Results}

In this section  we evaluate the packet outage probability  of the proposed \gls{bcfsaw}  against the range of feedback channel reliability metrics $p$ and compare it with  benchmark approaches introduced in \secref{Sec:Problem}.  All the results are presented for the case of $M = 4$ for repeated Monte Carlo analysis  where $\pe{} = 0.1$ and the combining gain is modeled by ${g}$  in $\pe{m} = (\pe{m-1})^{{g}}$ for $m>1$. For the case of \gls{arq} operation, combining gain is set to $g = 1$ while for the case of \gls{harq} operation we assume $g = 1.2$. The retransmission protocols are allowed to transmit only one packet (either initial transmission or retransmission) per transmit occasion. We further adopt the assumption of an error-free \gls{ndi} detection at the receiver node to solely  focus on the effects of unreliable feedback channel. 
Performance of the proposed \gls{bcfsaw} is evaluated for different  number of required \gls{ack} observances, $L$.  We assume a simple  retransmission phase packet order where upon observing a composite \gls{nack} the last active packet is retransmitted   until an \gls{ack} is observed. Otherwise,  when transmit counter  reaches $M$ for the packet  retransmission phase order  switches to the next last active packet and repeats the same process until all active packets  reach $M$ transmission attempts or an \gls{ack} is observed.

\begin{figure}[]
\begin{center}
\psfrag{xlabel}[c][c][\scalevalueS]{p}
\psfrag{ylabel}[c][c][\scalevalueS]{$\Tbar$ [$\times$ RTT]}
\psfrag{XXXXXXXXXXXX1}[lc][lc][\scalevalueSS]{\gls{regsaw}}
\psfrag{x2}[lc][lc][\scalevalueSS]{\gls{bcf}, $L=2$}
\psfrag{x3}[lc][lc][\scalevalueSS]{\gls{bcf}, $L=4$}
\psfrag{x4}[lc][lc][\scalevalueSS]{\gls{Lack}, $L = 2$}
\psfrag{x5}[lc][lc][\scalevalueSS]{\gls{Lack}, $L = 4$}
\psfrag{x6}[lc][lc][\scalevalueSS]{\gls{Lrep}, $L=2$}
\psfrag{x7}[lc][lc][\scalevalueSS]{\gls{Lrep}, $L=4$}
\psfrag{XXXXXXXXX8}[lc][lc][\scalevalueSS]{Asym, $\qn = \min \{10^{-5},p\}$}
\psfrag{x9}[lc][lc][\scalevalueSS]{Asym, $\qn = \min \{10^{-4},p\}$}
\psfrag{x10}[lc][lc][\scalevalueSS]{\gls{bretx}}
\includegraphics[width=0.98\columnwidth,keepaspectratio]{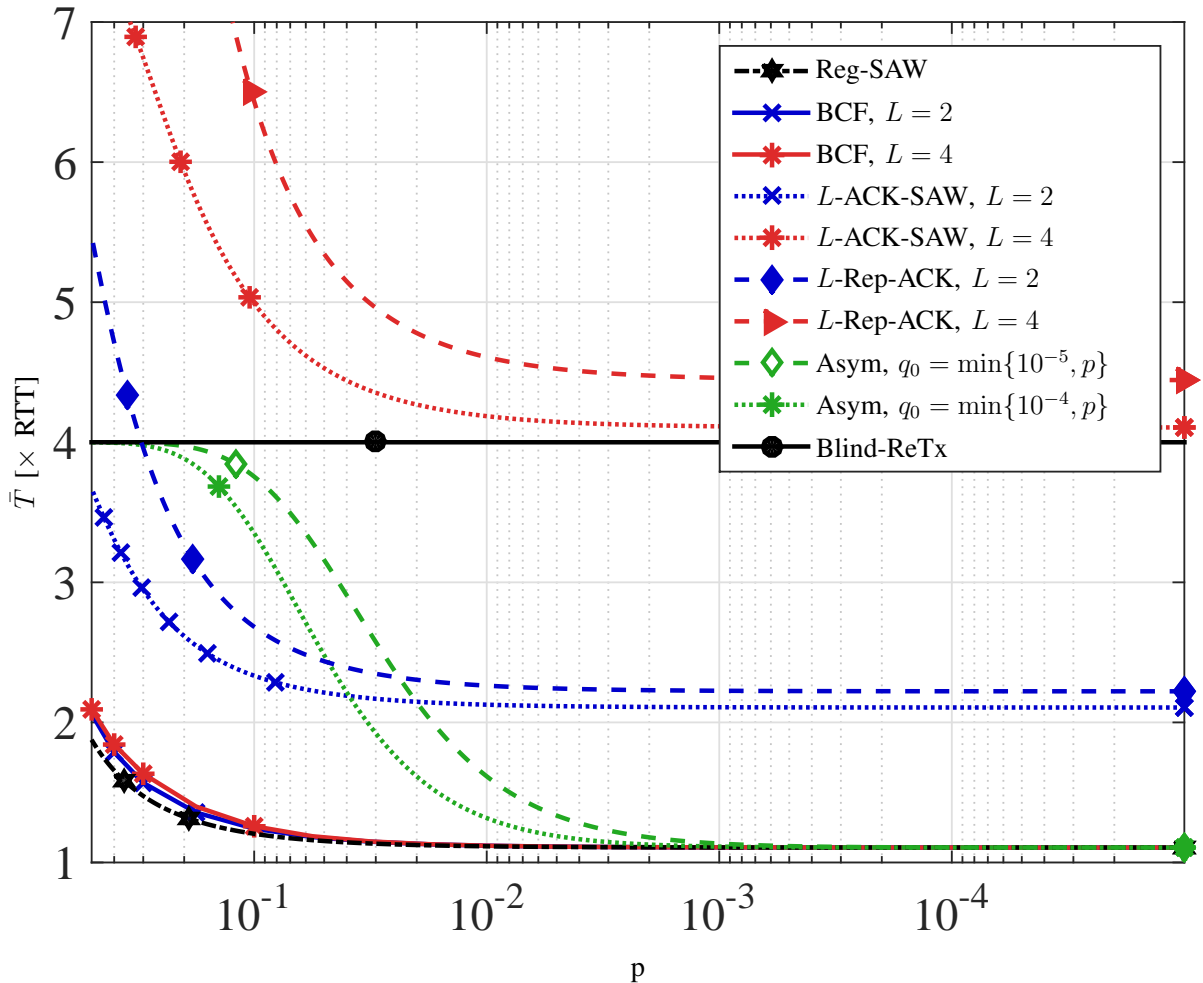}
\caption{Average number of feedback occasions utilized per packet packets $\Tbar$ for  $\pn = \pa = p$ and combining gain $g = 1.2$.}
\label{Fig:Tbar}
\end{center}
\vspace{-20pt}
\end{figure}

\begin{figure}[]
\begin{center}
\psfrag{xlabel}[c][c][\scalevalueS]{$p$}
\psfrag{ylabel}[c][c][\scalevalueS]{average number of channel use per packet, $\Nbar$}
\psfrag{XXXXXXXXX1}[lc][lc][\scalevalueSS]{\gls{regsaw}}
\psfrag{x2}[lc][lc][\scalevalueSS]{\gls{bcf}, $L=2$}
\psfrag{x3}[lc][lc][\scalevalueSS]{\gls{bcf}, $L=4$}
\psfrag{x4}[lc][lc][\scalevalueSS]{\gls{Lack}, $L = 2$}
\psfrag{x5}[lc][lc][\scalevalueSS]{\gls{Lack}, $L = 4$}
\psfrag{x6}[lc][lc][\scalevalueSS]{\gls{Lrep}, $L=2$}
\psfrag{x7}[lc][lc][\scalevalueSS]{\gls{Lrep}, $L=4$}
\psfrag{XXXXXXXXX8}[lc][lc][\scalevalueSS]{Asym, $\qn = \min \{10^{-5},p\}$}
\psfrag{x9}[lc][lc][\scalevalueSS]{Asym, $\qn = \min \{10^{-4},p\}$}
\psfrag{x10}[lc][lc][\scalevalueSS]{\gls{bretx}}
\includegraphics[width=0.99\columnwidth,keepaspectratio]{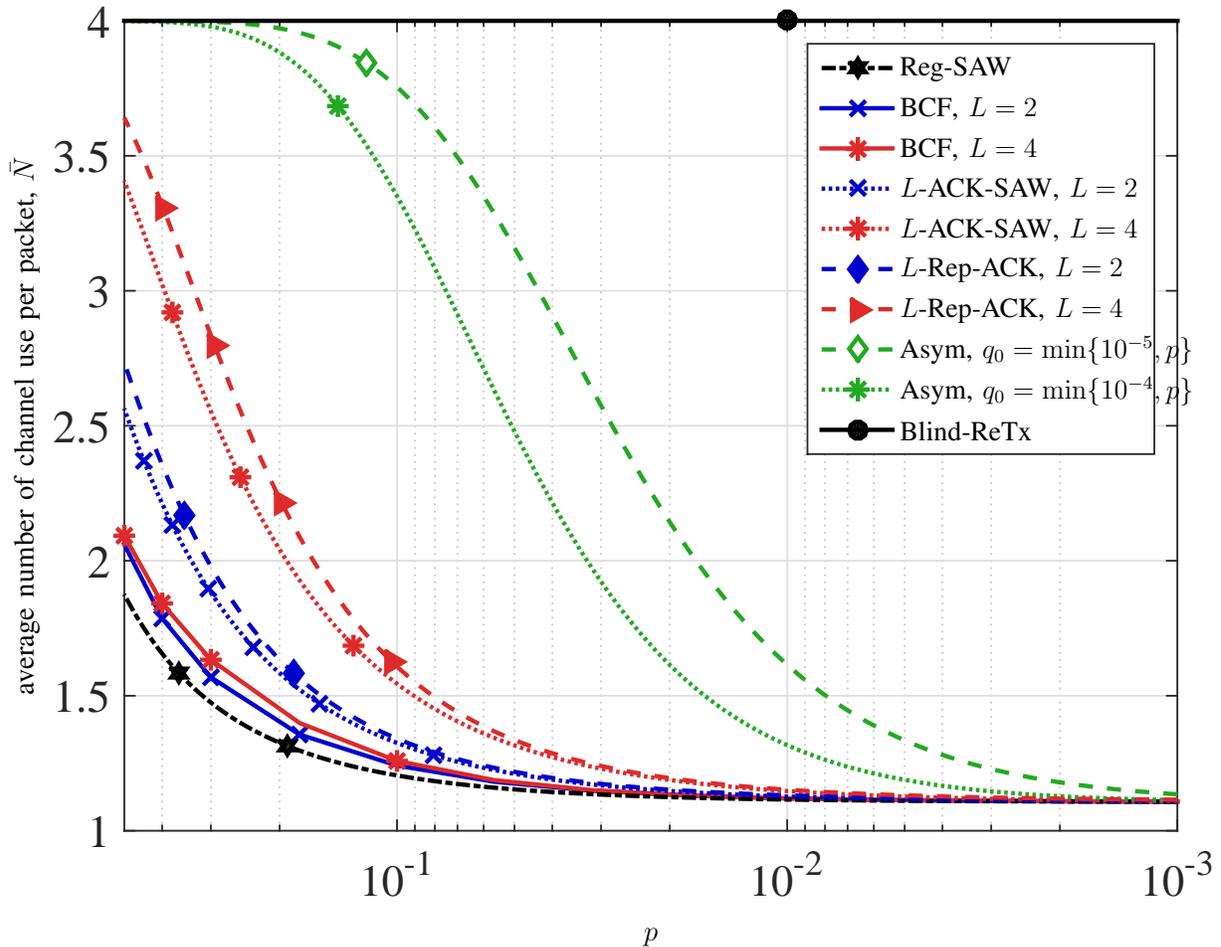}
\caption{Average  channel use per packet in number of transmit occasions for $M = 4$, $\pn = \pa = p$ and $\pe{} = 0.1$.}
\label{Fig:Nbar}
\end{center}
\end{figure}

Best case outage probability for  \gls{harq} and \gls{arq} operations reaches $\prod_m \pe{m}$ resulting in $4.285\tr{e}-6$ and $1\tr{e}-4$ limits respectively as shown in \figref{Fig:Outage}. The proposed \gls{bcfsaw}    reduces  outage probability by orders of magnitude e.g., for $L = 2$ and $L = 4$ as compared to \gls{regsaw} even in highly unreliable feedback cases while its outage performance is bounded by  \gls{Lack}. The latter performs more reliably because the $L$ observances of \gls{ack} are separately received for a given packet and a \gls{nack} feedback will trigger retransmission of the same packet. On the other hand, \gls{nack} observance in \gls{bcfsaw} may be followed by retransmission of a packet other than the failed packet incurring additional feedback occasions which may increase the false \gls{ack} rate. The better outage performance of  \gls{Lack} and \gls{Lrep} is thanks to the increased number of channel uses per packet $\Nbar$ as shown in  \figref{Fig:Nbar}.  However, the penalty paid for  improved reliability by the two latter approaches, as shown in \figref{Fig:Tbar}, is an increased average number of feedback occasions  per packet $\Tbar$ which is equivalent to the average experienced delay by higher layer application. $\Tbar$ increases almost linearly by increasing $L$ for those two methods resulting in a significantly higher penalty as compared to the the proposed \gls{bcfsaw}.

\begin{figure}[]
\begin{center}
\psfrag{xlabel}[c][c][\scalevalueS]{packet delivery latency}
\psfrag{ylabel}[c][c][\scalevalueS]{ccdf}
\psfrag{XXXXXXXXXXXX1}[lc][lc][\scalevalueSS]{\gls{bcf}, $L=4$}
\psfrag{x2}[lc][lc][\scalevalueSS]{\gls{Lack}, $L = 4$}
\psfrag{x3}[lc][lc][\scalevalueSS]{\gls{Lrep}, $L=4$}
\psfrag{x4}[lc][lc][\scalevalueSS]{\gls{Asym}, $\qn = 10^{-5}$}
\psfrag{x5}[lc][lc][\scalevalueSS]{\gls{bretx}}
\psfrag{xl01}[c][c][\scalevalueSS]{$t_0$}
\psfrag{xl02}[c][c][\scalevalueSS]{$t_1$}
\psfrag{xl03}[c][c][\scalevalueSS]{$t_2$}
\psfrag{xl04}[c][c][\scalevalueSS]{$t_3$}
\psfrag{xl05}[c][c][\scalevalueSS]{$t_4$}
\psfrag{xl06}[c][c][\scalevalueSS]{$t_5$}
\psfrag{xl07}[c][c][\scalevalueSS]{$t_6$}
\psfrag{xl08}[c][c][\scalevalueSS]{$t_7$}
\psfrag{xl09}[c][c][\scalevalueSS]{$t_8$}
\psfrag{xl10}[c][c][\scalevalueSS]{$t_9$}
\psfrag{xl11}[c][c][\scalevalueSS]{$t_{10}$}
\psfrag{xl12}[c][c][\scalevalueSS]{$t_{11}$}
\psfrag{xl13}[c][c][\scalevalueSS]{$t_{12}$}
\psfrag{xl14}[c][c][\scalevalueSS]{$t_{13}$}
\psfrag{xl15}[c][c][\scalevalueSS]{$t_{14}$}
\psfrag{xl16}[c][c][\scalevalueSS]{$t_{15}$}
\includegraphics[width=0.98\columnwidth,keepaspectratio]{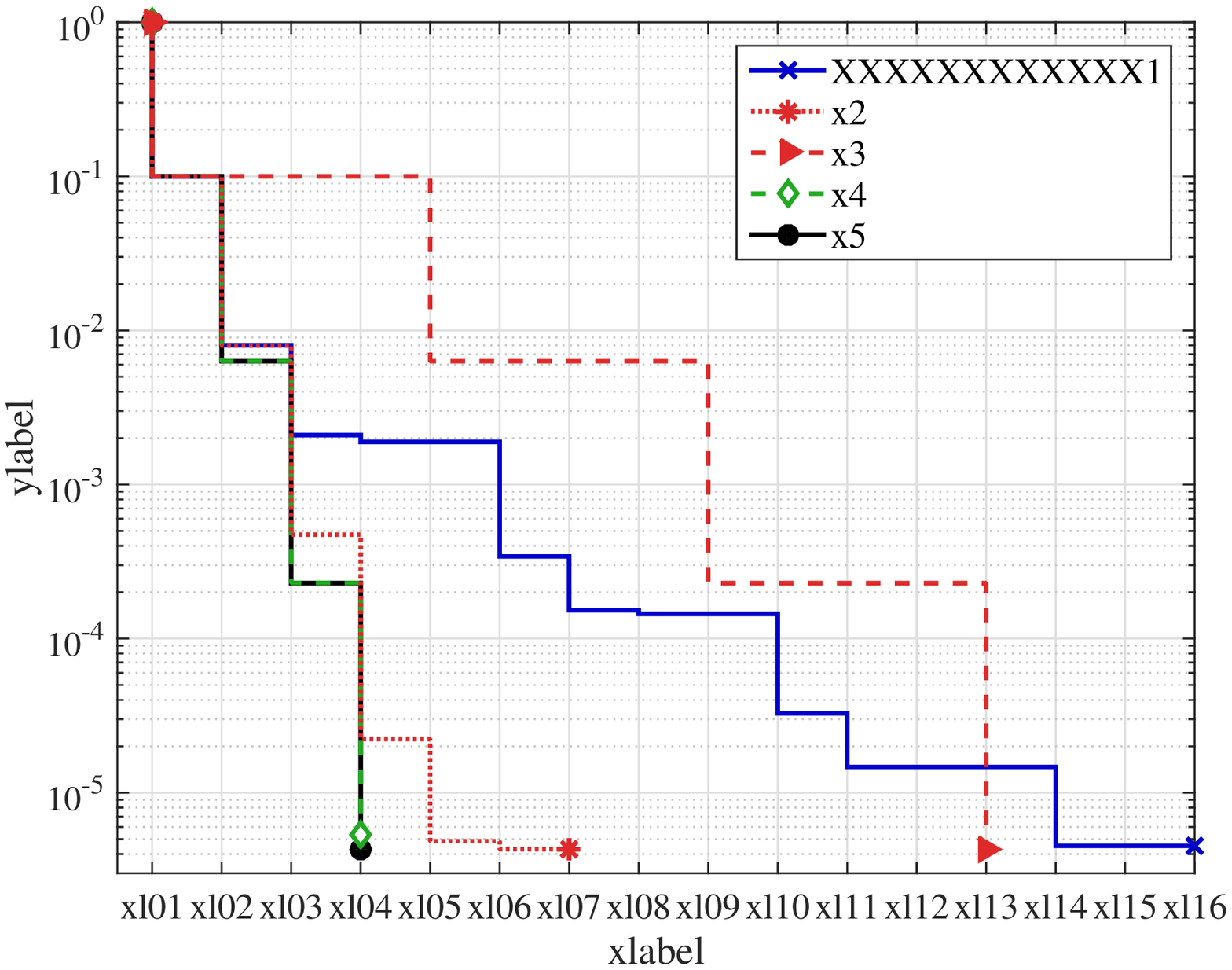}
\caption{\gls{ccdf} of the packet delivery latency at $p = 2\tr{e}-2$ with combining gain $g = 1.2$, where $t_{k} = \tr{TTI}+ k*\tr{RTT}$.}
\label{Fig:Latency}
\end{center}
\vspace{-15pt}
\end{figure}

In \figref{Fig:Latency},   \gls{ccdf} of packet delivery latency is shown for all the approaches achieving $\Pout \leq 10^{-5}$ in $R5$ from \figref{Fig:Outage}. The best case latency performance is achieved using \gls{bretx} and \gls{Asym} with $\qn = 10^{-5}$. While \gls{Lack} with $L = 4$  provides a better latency statistic than \gls{bcfsaw}, it fails in the average delay experienced by the higher layer application. This increases number of feedback reporting per packet transmit occasion roughly by $L$. On the other hand,  \gls{bcfsaw} uses roughly the same average number of feedback reporting per packet as compared to \gls{regsaw} while providing higher reliability. 
 By comparison of the presented numerical results in different ultra-reliability operation regions the following observations can be made.
\begin{itemize}
\item	In fairly reliable feedback channel conditions, e.g., lower $p$ regime in region $R4$, \gls{Asym} provides   high reliability with low  $\Tbar$ which makes it a viable choice for only when $p$ is ideally low.
\item	In unreliable feedback conditions, e.g., region $R5$ and higher $p$ regime in $R4$, \gls{bcfsaw} is the most viable solution for ultra-reliable communication with low energy and low cost receiver type  where low $\Tbar$ is required. For use cases  with low latency requirement,  \gls{Lack} approach performs better  if $L < M$ however, it requires relaxed limitations on receiver node energy consumption and assuming low traffic channel (i.e., where high $\Tbar$ is tolerated). Otherwise, for $L \geq M$, \gls{bretx} performs more efficiently than \gls{Lack} with less energy consumption for feedback and guaranteed ultra reliability.
\item	In extremely unreliable feedback channel conditions, e.g., region $R6$, \gls{bretx} is the better choice over \gls{Asym}, providing similar performance without the need for  feedback channel.
\end{itemize}

%
%
%

%

\section{Conclusions}
\label{Sec:Conclusions}

We proposed   a new method of acknowledging packet delivery for  unreliable feedback channel conditions. The proposed method, dubbed  \gls{bcfsaw}, relies on backwards composite  acknowledgement and provides the  retransmission protocols  with  configurable ultra-reliability. It further provides the scheduler of the wireless system  with new degrees of freedom  to configure the communication link in order to meet the desirable  reliability requirement  even in highly-unreliable feedback channel conditions. The presented numerical analysis show orders of magnitude increase in reliability of the retransmission protocols over the practical range of target block error rate only at the cost of a negligible increase in average  experienced packet delay.   System-level performance analysis of the proposed method in more realistic  multi-user communication systems with time-varying channel conditions will be studied as future work.


\bibliographystyle{IEEEtran}
\bibliography{IEEEabrv,references_all}{}

\begin{thebibliography}{1}
\providecommand{\url}[1]{#1}
\csname url@samestyle\endcsname
\providecommand{\newblock}{\relax}
\providecommand{\bibinfo}[2]{#2}
\providecommand{\BIBentrySTDinterwordspacing}{\spaceskip=0pt\relax}
\providecommand{\BIBentryALTinterwordstretchfactor}{4}
\providecommand{\BIBentryALTinterwordspacing}{\spaceskip=\fontdimen2\font plus
\BIBentryALTinterwordstretchfactor\fontdimen3\font minus
  \fontdimen4\font\relax}
\providecommand{\BIBforeignlanguage}[2]{{%
\expandafter\ifx\csname l@#1\endcsname\relax
\typeout{** WARNING: IEEEtran.bst: No hyphenation pattern has been}%
\typeout{** loaded for the language `#1'. Using the pattern for}%
\typeout{** the default language instead.}%
\else
\language=\csname l@#1\endcsname
\fi
#2}}
\providecommand{\BIBdecl}{\relax}
\BIBdecl

\bibitem{Lin:1984}
S.~Lin \emph{et~al.}, ``Automatic-repeat-request error-control schemes,''
  \emph{Comm. Mag.}, vol.~22, no.~12, pp. 5--17, Dec. 1984.

\bibitem{3gpp38802}
``{3GPP TR} 38.802; study on new radio access technology physical layer
  aspects,'' 3GPP, Tech. Rep., Mar. 2017.

\bibitem{3gpp36212}
``{3GPP TS} 36.212; evolved universal terrestrial radio access {(E-UTRA)};
  multiplexing and channel coding,'' 3GPP, Tech. Spec., Dec. 2016, release 10.

\bibitem{3gpp36213}
``{3GPP TS} 36.213; evolved universal terrestrial radio access {(E-UTRA)};
  physical layer procedures,'' 3GPP, Tech. Spec., Jan. 2015, release 10.

\bibitem{3gpp36877}
``{3GPP TR} 36.877; {LTE} device to device {(D2D)} proximity services
  {(ProSe)},'' 3GPP, Tech. Rep., Mar. 2015, release 12.

\bibitem{khosravirad2016overview}
S.~R. Khosravirad \emph{et~al.}, ``Enhanced {HARQ} design for {5G} wide area
  technology,'' in \emph{IEEE Vehicular Technology Conference (VTC'16)}, May
  2016.

\bibitem{Goldsmith:2005}
A.~Goldsmith, \emph{Wireless Communications}.\hskip 1em plus 0.5em minus
  0.4em\relax Cambridge University Press, Aug. 2007.

\end{thebibliography}

\end{document}